\documentclass[12pt, preprint]{aastex}
\newcommand{\kms}{\,km~s$^{-1}$}

\def\spose#1{\hbox to 0pt{#1\hss}}
\def\simlt{\mathrel{\spose{\lower 3pt\hbox{$\mathchar"218$}}
     \raise 2.0pt\hbox{$\mathchar"13C$}}}
\def\simgt{\mathrel{\spose{\lower 3pt\hbox{$\mathchar"218$}}
     \raise 2.0pt\hbox{$\mathchar"13E$}}}
\slugcomment{Resubmitted to ApJ on August 18, 2006}
\usepackage{epsf}
\font\smcap=cmcsc10

\def\arcsec{$''$}

\def\ki{K\,{\smcap i}}
\def\nai{Na\,{\smcap i}}
\def\caii{Ca\,{\smcap ii}}

\def\ddo{DDO51}

\def\vio{$(V-I)_0$}
\def\ivi{($I,\,V-I$)}
\def\xcmd{$X_{\rm CMD}$}
\def\ycmd{$Y_{\rm CMD}$}
\def\fehp{$\rm[Fe/H]_{phot}$}
\def\fehs{$\rm[Fe/H]_{spec}$}
\def\pg{$P_{\rm giant}$}
\def\pd{$P_{\rm dwarf}$}
\shorttitle{Isolating Red Giants in M31's Outer Halo}
\shortauthors{Gilbert et~al.}

\begin{document}

\title{A New Method for Isolating M31 Red Giant Stars: The Discovery of Stars
out to a Radial Distance of 165 Kiloparsecs}

\author{
Karoline~M.~Gilbert\altaffilmark{1},
Puragra~Guhathakurta\altaffilmark{1},
Jasonjot~S.~Kalirai\altaffilmark{1,2},
R.~Michael~Rich\altaffilmark{3},
Steven~R.~Majewski\altaffilmark{4},
James~C.~Ostheimer\altaffilmark{4},
David~B.~Reitzel\altaffilmark{3}, 
A.~Javier~Cenarro\altaffilmark{1,5},
Michael~C.~Cooper\altaffilmark{6},
Carynn~Luine\altaffilmark{1}, and
Richard~J.~Patterson\altaffilmark{4}
}

\email{
kgilbert@astro.ucsc.edu,
raja@ucolick.org,
jkalirai@ucolick.org,
rmr@astro.ucla.edu,
srm4n@virginia.edu,
jostheim@alumni.virginia.edu,
reitzel@astro.ucla.edu,
cen@astrax.fis.ucm.es,
cooper@astron.berkeley.edu,
kierrica@hotmail.com,
rjp0i@virginia.edu
}

\altaffiltext{1}{UCO/Lick Observatory, Department of Astronomy \&
Astrophysics, University of California Santa Cruz, 1156 High Street, Santa
Cruz, California 95064, USA.}
\altaffiltext{2}{Hubble Fellow.}
\altaffiltext{3}{Department of Physics \& Astronomy, Knudsen Hall, University
of California, Los Angeles, California 90095, USA.}
\altaffiltext{4}{Department of Astronomy, University of Virginia, PO
Box~3818, Charlottesville, Virginia 22903, USA.}
\altaffiltext{5}{Present address: Departamento de Astrof\'isica, Universidad
Complutense de Madrid, E-28040 Madrid, Spain.}
\altaffiltext{6}{Department of Astronomy, Campbell Hall, University of
California, Berkeley, California 94720, USA.}

\setcounter{footnote}{6}

%%%%%%%%%%%%%%%%%%%%%%%%%%%%%%%%%%%%%%%%%%%%%%%%%%%%%%%%%%%%%%%%%%%%%%%%%%%%%
\begin{abstract}
We present a method for isolating a clean sample of red giant branch stars in
the outer regions of the Andromeda spiral galaxy (M31).  Our study is based
on an ongoing spectroscopic survey using the DEIMOS instrument on the Keck~II
10-m telescope\footnote{Data presented herein were obtained at the W.\ M.\
Keck Observatory, which is operated as a scientific partnership among the
California Institute of Technology, the University of California and the
National Aeronautics and Space Administration.  The Observatory was made
possible by the generous financial support of the W.\ M.\ Keck Foundation.}.
The survey aims to study the kinematics, global structure, substructure, and
metallicity of M31's halo.  Although most of our spectroscopic targets were
photometrically screened to reject foreground Milky Way dwarf star
contaminants, the latter class of objects still constitutes a substantial
fraction of the observed spectra in the sparse outer halo.  Our
likelihood-based method for isolating M31 red giants uses five criteria:
(1)~radial velocity, (2)~photometry in the intermediate-width \ddo\ band to
measure the strength of the MgH/Mg\,$b$ absorption features, (3)~strength of
the \nai\ 8190~\AA\ absorption line doublet, (4)~location within an \ivi\
color-magnitude diagram, and (5)~comparison of photometric (CMD-based) versus
spectroscopic (\caii\ 8500~\AA\ triplet-based) metallicity estimates.  We
also discuss other giant/dwarf separation criteria that might be useful in
future analyses: the strength of the \ki\ absorption lines at 7665 and
7699~\AA\ and the TiO bands at 7100, 7600, and 8500~\AA.  Training sets
consisting of definite M31 red giants and Galactic dwarf stars are used to
derive empirical probability distribution functions for each diagnostic.
These functions are used to calculate the likelihood that a given star is a
red giant branch star in M31 versus a Milky Way dwarf star.  By applying this
diagnostic method to our spectroscopic data set, we isolate 40~M31 red giants
beyond a projected distance of $R=60$~kpc from the galaxy's center, including
three~red giants out at $R\sim165$~kpc.  The ability to identify individual
M31 red giant stars gives us an unprecedented level of sensitivity in
studying the properties of the galaxy's outer halo.
\end{abstract}


% \setcounter{footnote}{0}

%%%%%%%%%%%%%%%%%%%%%%%%%%%%%%%%%%%%%%%%%%%%%%%%%%%%%%%%%%%%%%%%%%%%%%%%%
\section{Introduction}\label{sec:intro}
%%%%%%%%%%%%%%%%%%%%%%%%%%%%%%%%%%%%%%%%%%%%%%%%%%%%%%%%%%%%%%%%%%%%%%%%%%%%%

Hierarchical galaxy formation theories propose that galaxies are built up
through the accretion of smaller systems \citep{sea78,whi78}.  Lately there
has been significant progress in the investigation of galactic halos,
especially in studies focusing on the contribution of tidal debris from
disrupted satellites.  Computational studies, using numerical simulations and
semi-analytic modelling, have made great strides in understanding the
properties of halos built up from tidal debris
\citep*[e.g.,][]{joh96,joh98,hel99,hel00,bul01,bul05}.  Recent observational
studies have led to the discovery of large tidal streams in the Milky Way
(MW) halo such as the Magellanic Stream \citep*{mat74}, the Sagittarius
stream \citep*{iba94,maj03,new03}, and the Monoceros stream
\citep{yan03,roc03}.  These observations are adding new insight to the study
of halo formation.

Studies of tidal structures in the halo of the Galaxy present inherent
difficulties, as they generally require wide-field surveys given their large
angular extent when viewed from our vantage point within the disk.  The
Andromeda spiral galaxy (M31) presents a unique opportunity to study a galaxy
like our own from an external perspective.  In addition, it has the advantage
of having a high disk inclination angle of $78^\circ$ \citep{deV58,wal88},
and at a distance of 783~kpc \citep{sta98,hol98} it is close enough to allow
detailed photometric studies of individual stars
\citep*[e.g.,][]{mou86,dur01,dur04,bel03}, even faint main-sequence turnoff
stars \citep{bro03,bro06}, as well as spectroscopy of individual red giant
branch (RGB) stars \citep*{rei02,rei04,iba04,iba05,p1,kal06b,cha06}.

As in the MW, recent studies of M31 have uncovered evidence of past and
present accretion events.  Star-count maps covering a large area of M31 have
revealed a `giant southern stream' and many other signs of disturbance in the
spheroid and outer disk \citep{iba01,fer02}.  Evidence of tidal disruption in
M31's closest satellite galaxies M32 and NGC~205 has been presented by
\citet{cho02}.

\setcounter{footnote}{7}

In order to investigate the substructure of the outer halo through kinematics
and metallicity, we have undertaken a spectroscopic survey of the M31 halo
with the Deep Imaging Multi-Object Spectrograph \citep*[DEIMOS;][]{fab03} on
the Keck~II Telescope.  The first results from this survey focus on the giant
southern stream discovered by \citet{iba01}, and are presented by \citet{p1},
\citet{fon06}, and \citet{kal06b}.

The ability to take spectra of individual stars in the halo of M31 opens up
many exciting avenues for studying its kinematics, structure, and metallicity
\citep{p1,kal06b}.  Radial velocities are the standard
method for discriminating between RGB stars in M31 and dwarf stars in the MW.
However, the radial velocity distributions of these two populations overlap.
Although modelling the distribution of radial velocities can provide a
{\it statistical\/} separation of M31 RGB versus Galactic dwarf stars, a more
stringent method for differentiating the two populations is needed to mark
individual stars as M31 RGB or MW dwarf stars.  This is vital for studying
the far outer regions of the halo, where our survey is expected to find only
a handful (if any) RGB stars per DEIMOS pointing, and will give us the
sensitivity needed to probe M31's halo to extremely low surface brightness 
levels.

In this paper, we describe a method that uses a combination of spectroscopic
and photometric information to determine whether a star is an M31 RGB star or
a foreground MW dwarf star.  This allows us to isolate a clean sample of RGB
stars in M31.  Details of the observations and data reduction are summarized
in \S\,\ref{sec:data}.  The selection of training set stars, the different
spectroscopic and photometric diagnostics, probability distribution
functions, and the application of a likelihood-based screening technique are
described in \S\,\ref{sec:method}.  In \S\,\ref{sec:disc}, the robustness of
the diagnostic method is tested and it is compared to other methods used to
study RGB stars in M31.  The main points of the paper are summarized in
\S\,\ref{sec:concl}. 

%%%%%%%%%%%%%%%%%%%%%%%%%%%%%%%%%%%%%%%%%%%%%%%%%%%%%%%%%%%%%%%%%%%%%%%%%
\section{Data}\label{sec:data}
%%%%%%%%%%%%%%%%%%%%%%%%%%%%%%%%%%%%%%%%%%%%%%%%%%%%%%%%%%%%%%%%%%%%%%%%%%%%%
%%%%%%%%%%%%%%%%%%%%%%%%%%%%%%%%%%%%%%%%%%%%%%%%%%%%%%%%%%%%%%%%%%%%%%%%%%
\subsection{Slitmask Design and Observations}\label{sec:slit_design}
%%%%%%%%%%%%%%%%%%%%%%%%%%%%%%%%%%%%%%%%%%%%%%%%%%%%%%%%%%%%%%%%%%%%%%%%%%

Objects were selected for Keck/DEIMOS spectroscopy using the M31
photometry/astrometry catalogs of \citet{ost02}.  The catalogs were derived
from images taken with the Mosaic camera on the Kitt Peak National
Observatory (KPNO)\footnote{Kitt Peak National Observatory of the National
Optical Astronomy Observatory is operated by the Association of Universities
for Research in Astronomy, Inc., under cooperative agreement with the
National Science Foundation} 4-m telescope in the Washington System $M$ and
$T_2$ bands and the intermediate-width \ddo\ band.  The \ddo\ filter is
centered at a wavelength of 5150~\AA\ with a width of about 100~\AA.  It
includes the surface-gravity sensitive Mg\,$b$ and MgH stellar absorption
features, which are strong in dwarf stars but weak in RGB stars.  Photometric
transformation relations from \citet{maj00} were used to derive
Johnson-Cousins $V$ and $I$ magnitudes from the $M$ and $T_2$ magnitudes.

Due to the sparseness of RGB stars the outer halo of M31, objects were
photometrically pre-selected in order to maximize the efficiency of the
spectroscopic observations.  Each object was assigned a \ddo\ parameter, an
estimate of the probability of its being an RGB star based on its position in
the ($M-DDO51$) versus ($M-T_2$) color-color diagram \citep{pal03}.  Objects
with point-like morphology (based on the DAOPHOT-based morphological
parameters {\tt chi} and {\tt sharp}) and large values of the \ddo\ parameter
were given the highest priority during the design of the DEIMOS slitmask.
Background galaxies tend to get assigned high values of the \ddo\ parameter
(either because their Mg\,$b$/MgH features are redshifted out of the \ddo\
band or because they are dominated by light from RGB stars), but all but the
most compact ones fail the {\tt chi} and {\tt sharp} criteria.  \citet{p1}
presents details of the spectroscopic target selection and slitmask design
for the outer halo fields in our survey.   

In addition to outer halo fields, we have obtained Keck/DEIMOS observations
of three fields relatively close to the center of M31 that have recently been
imaged to below the main sequence turnoff with the Advanced Camera for
Surveys on the {\it Hubble Space Telescope\/} \citep{bro03,bro06}.  These
fields are on the southeastern minor axis (H11), giant southern stream
(H13s), and northeastern major axis (H13d).  We do not have \ddo/Washington
system photometry for these fields so we did not use the $P_{\rm giant}$
criterion for spectroscopic target selection.  The surface density of M31 RGB
stars is very high in these inner fields, leading to a large RGB to dwarf ratio even without preselection of RGB star candidates.
Photometry for the H11 and H13s fields comes from MegaCam images taken with
the 3.6-m Canada-France-Hawaii Telescope (CFHT)\footnote{MegaPrime/MegaCam is
a joint project of CFHT and CEA/DAPNIA, at the Canada-France-Hawaii Telescope
which is operated by the National Research Council of Canada, the Institut
National des Science de l'Univers of the Centre National de la Recherche
Scientifique of France, and the University of Hawaii.} in the $g'$ and $i'$
bands.  Object detection and photometry for these fields was done using
SExtractor \citep{ber96} and the instrumental magnitudes were transformed to
the Johnson-Cousins $V$ and $I$ bands \citep{kal06b}.  Photometry for the
H13d fields comes from short Keck/DEIMOS imaging exposures in the $V$ and $I$
bands (Reitzel et~al.\ 2006, in preparation).  The lists for slitmask design
in the H11, H13s, and H13d fields were based on $I$ magnitude and the
SExtractor morphological criteron {\tt stellarity}.

\begin{figure}
\centerline{\epsfxsize=6in \epsfysize=6in
\epsfbox{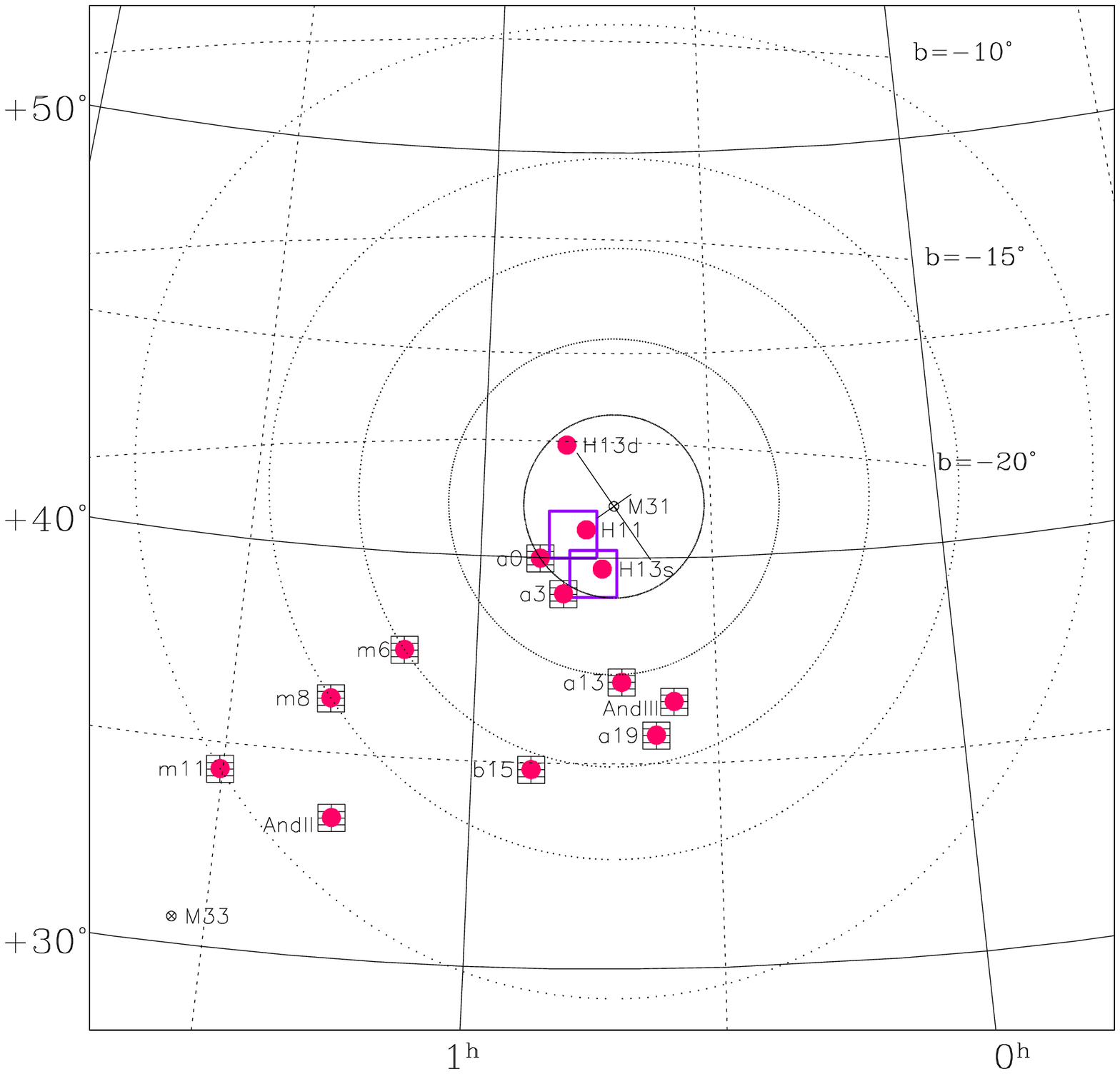}}
\figcaption[fig01.eps]{\label{fig:map}Sky position (J2000.0 RA/DEC) of the
fields observed with DEIMOS spectroscopic slitmasks (solid circles).  The
small gridded squares indicate the full area ($35'\times35'$) of the fields
from the KPNO 4-m/Mosaic imaging survey of \citet{ost02}.  The resulting
photometric $DDO51$-, $M$-, and $T_2$-band catalog was used to select
spectroscopic targets for the DEIMOS slitmasks, each of which covers an area
of roughly $16'\times4'$.  The majority of the fields in our $DDO51$ survey
lie close to the south-eastern minor axis of M31 and extend out to a
projected distance from M31 of $R=163$~kpc (field~m11).  The large bold open
squares represent two of our CFHT MegaCam $g'$- and $i'$-band pointings,
used to select spectroscopic targets for fields H11 and H13s.  The pair of
thin solid diagonal lines indicates the approximate extent of the visible
disk of M31.  The location of M33 is marked near the lower left.  The
concentric circles correspond to projected distances from M31's center of
30, 60, 90, 120, and 165~kpc.  A few lines of Galactic latitude are also
marked.
}
\end{figure}

Spectroscopic observations of DEIMOS multislit masks were obtained in
fall~2002, 2003, 2004, and 2005 (Table~\ref{tab:masks}; Fig.~\ref{fig:map})
using the Keck~II telescope and the DEIMOS instrument with the
1200~line~mm$^{-1}$ grating.  This grating yields a dispersion of
$\rm0.33~\AA$~pix$^{-1}$ and the spatial scale is 0$\farcs$12~pix$^{-1}$.
For masks observed in fall~2002, the central wavelength setting was
$\rm\lambda8550~\AA$, and the spectra cover the range
$\lambda\lambda7200$--$\rm9900~\AA$.  This setting had the unfortunate effect
that one of the lines of the \caii\ triplet occasionally landed in the
inter-CCD gap in DEIMOS.  To avoid this, and to extend the spectral coverage
to shorter wavelengths to include the $\rm\lambda7100~\AA$ TiO and
$\rm\lambda6563~\AA$ H$\alpha$ features, masks observed in fall~2003, 2004,
and 2005 had the central wavelength set to $\rm\lambda7800~\AA$, yielding the
spectral coverage $\lambda\lambda6450$--$\rm9150~\AA$.  Slits had a width of
1\arcsec, which subtends 4.8~pix.  The spectral resolution is slightly better
than this: $\rm3.8~pix=1.26~\AA$ for the typical seeing conditions
($0\farcs8$ FWHM).  This resolution corresponds to 44~km~s$^{-1}$ at the
\caii\ triplet.  Our actual radial velocity measurement error is
significantly smaller than this (see \S\,\ref{sec:error})---the centroiding
accuracy depends on the signal-to-noise (S/N) ratio and is typically much
smaller than the width of the line \citep{ton79}. 

%%%%%%%%%%%%%%%%%%%%%%%%%%%%%%%%%%%%%%%%%%%%%%%%%%%%%%%%%%%%%%%%%%%%%%%%%%%%%
\begin{table}[ht!]
\begin{center}
\caption{Details of spectroscopic observations and basic results.}
\vskip 0.3cm
\begin{tabular}{rrrrrllrr}
\hline
\hline
\multicolumn{1}{c}{Mask}           & \multicolumn{2}{c}{Pointing center:} &
\multicolumn{1}{c}{PA} & \multicolumn{1}{c}{\#\ Sci.}  &
\multicolumn{1}{c}{Date of} & 
\multicolumn{1}{c}{Number of} \\ 
            & \multicolumn{1}{c}{$\alpha_{\rm J2000}$} &
\multicolumn{1}{c}{$\delta_{\rm J2000}$} &\multicolumn{1}{c}{($^\circ$E of N)}
&\multicolumn{1}{c}{targets\tablenotemark{a}} &\multicolumn{1}{c}{Obs. (UT)}
 &\multicolumn{1}{c}{$Q=-2$/1/2/3/4} \\
      & \multicolumn{1}{c}{($\rm^h$:$\rm^m$:$\rm^s$)} &
\multicolumn{1}{c}{($^\circ$:$'$:$''$)} & & & & 
\multicolumn{1}{c}{cases} \\
\hline
H11\_1      & 00:46:21.02 & +40:41:31.3 & 21.0    & 139 & 2004 Sep 20 &
40/01/29/10/59\tablenotemark{b} \\
H11\_2      & 00:46:21.02 & +40:41:31.3 & $-21.0$ & 138 & 2004 Sep 20 & 
03/00/55/10/70 \\
H13s\_1     & 0:44:14.76  & +39:44:18.2 & 21.0    & 134 & 2004 Sep 20 & 
05/01/58/16/54 \\
H13s\_2     & 0:44:14.76  & +39:44:18.2 & $-21.0$ & 138 & 2004 Sep 20 &
44/00/33/11/50\tablenotemark{b} \\ 
H13d\_1     & 0:49:04.80  & +42:45:22.6 & 27.8    & 146 & 2004 Sep 20 & 
01/01/41/22/81 \\
H13d\_2     & 0:49:04.80  & +42:45:22.6 & 27.8    & 144 & 2004 Sep 20 & 
00/01/34/20/89 \\
a3\_1       & 00:48:21.16 & +39:02:39.2 & 64.2    & 85  & 2002 Aug 16 &
08/22/14/15/26 \\
a3\_2       & 00:47:47.24 & +39:05:56.3 & 178.2   & 81  & 2002 Oct 11 &
07/19/14/09/32 \\
a3\_3       & 00:48:23.17 & +39:12:38.5 & 270.0   & 83  & 2003 Oct 26 &
02/00/22/11/48 \\
a0\_1       & 00:51:51.32 & +39:50:21.4 & $-17.9$ & 89  & 2002 Aug 16 &
05/00/35/11/38 \\
a0\_2       & 00:51:29.59 & +39:44:00.8 & 90.0    & 89  & 2002 Oct 12 & 
02/00/42/15/30 \\
a0\_3       & 00:51:50.46 & +40:07:00.9 & 0.0     & 90  & 2004 Jun 17 & 
07/01/24/10/47 \\
a13\_1      & 00:42:58.34 & +36:59:19.3 & 0.0     & 80  & 2003 Sep 30 & 
07/04/08/05/56 \\
a13\_2      & 00:41:28.27 & +36:50:19.2 & 0.0     & 71  & 2003 Sep 30 & 
03/01/05/02/60 \\
a19\_1      & 00:38:16.05 & +35:28:07.2 & $-90.0$ & 71  & 2005 Aug 29 & 
03/00/04/06/58 \\
m6\_1       & 01:09:51.75 & +37:46:59.8 & 0.0     & 75  & 2003 Oct 1  & 
04/01/05/04/61 \\
m6\_2       & 01:08:36.22 & +37:28:59.6 & 0.0     & 72  & 2005 Jun 9  &  
05/01/05/12/49 \\
b15\_1      & 00:53:23.63 & +34:37:16.0 & $-90.0$ & 65  & 2005 Sep 7  & 
07/00/07/08/43 \\
b15\_3      & 00:53:37.77 & +34:50:04.1 & $-90.0$ & 74  & 2005 Sep 7  &  
05/02/09/04/54 \\
m8\_1       & 01:18:11.56 & +36:16:24.9 & 0.0     & 56  & 2005 Jul 7  & 
03/00/03/02/48 \\
m8\_2       & 01:18:35.87 & +36:14:30.9 & 0.0     & 59  & 2005 Jul 7  & 
03/00/04/07/45 \\
m11\_1      & 01:29:34.44 & +34:13:45.4 & 0.0     & 72  & 2003 Sep 30 & 
05/03/10/07/47 \\ 
m11\_2      & 01:29:34.35 & +34:27:45.5 & 0.0     & 68  & 2003 Oct 1  & 
04/03/17/02/42 \\
m11\_3      & 01:30:01.53 & +34:13:45.4 & 0.0     & 80  & 2005 Jul 8  & 
01/03/11/12/53 \\
m11\_4      & 01:30:37.33 & +34:13:27.4 & 0.0     & 75  & 2005 Jul 8  & 
02/02/10/10/51 \\
\hline
\end{tabular}
\tablenotetext{a}{A number of stars were observed on two different masks
(\S\,\ref{sec:error}).}
\tablenotetext{b}{The high number of failed targets in these masks reflects
incomplete data reduction for these masks.  Our success rate for these masks
will increase once the remaining slits are reduced.}
\label{tab:masks}
\end{center}
\end{table}
%%%%%%%%%%%%%%%%%%%%%%%%%%%%%%%%%%%%%%%%%%%%%%%%%%%%%%%%%%%%%%%%%%%%%%%%%%%%%

%%%%%%%%%%%%%%%%%%%%%%%%%%%%%%%%%%%%%%%%%%%%%%%%%%%%%%%%%%%%%%%%%%%%%%%%%%%%%
\subsection{Spectroscopic Data Reduction}\label{sec:dataredux}
%%%%%%%%%%%%%%%%%%%%%%%%%%%%%%%%%%%%%%%%%%%%%%%%%%%%%%%%%%%%%%%%%%%%%%%%%%%%%
Spectra were reduced using the {\tt spec2d} and {\tt spec1d} software
developed by the DEEP2 team at the University of California,
Berkeley\footnote{{\tt
http://astron.berkeley.edu/$\sim$cooper/deep/spec1d/primer.html}, 
\newline
{\tt http://astron.berkeley.edu/$\sim$cooper/deep/spec2d/primer.html}}.  
Spectra were flat fielded and corrected for fringing, and a wavelength
solution was determined from arc lamp exposures.  Each slitlet was sky
subtracted, and individual exposures were averaged using cosmic-ray rejection
and inverse-variance weighting.  Extracted one-dimensional spectra were
cross-correlated against stellar and galaxy templates spanning a range of
spectral types to determine the redshift.  The ten best cross-correlation
matches were reported by the {\tt spec1d} pipeline.  These matches were
evaluated by eye using the visual inspection software {\bf zspec} [developed
by D.~Madgwick and one of us (M.C.C.) for the DEEP2 survey] and the best
one identified.  In rare cases, the correct redshift did not appear among
the top ten cross-correlation matches in spite of the presence of one or more
visually identifiable spectral features [such as the \caii\ triplet, \nai\
doublet (dwarf stars), and $\lambda7100$~\AA\ TiO band]; the redshift had to
be determined manually in these cases.  Each spectrum was assigned a quality
code: spectra with secure redshifts was assigned $Q=4$ (at least two good
spectral lines) or $Q=3$ (one good line and one marginal line).  Objects for
which the spectra did not yield a reliable redshift were assigned $Q=2$
(visible continuum with low S/N or no lines), $Q=1$ (very weak and/or
undetectable continuum), or $Q=-2$ (catastrophic instrument failure).  We
make no distinction between $Q=4$ and $Q=3$ spectra in our analysis.
Table~\ref{tab:masks} lists the number of targets per quality bin for each
mask.  For more details about the reduction of the survey data see
\citet{p1}.

Spectra which have secure redshift measurements ($Q=4$ or 3) are shifted to
the rest frame and rebinned onto a common, uniform (linear) wavelength grid.
The continuum level of each spectrum is normalized to unity and the S/N
computed in the region of the \caii\ triplet (both the overall S/N and the
S/N between strong atmospheric emission lines).  The S/N calculation is based
on the assumption of Poisson errors as reported by {\tt spec2d} for each
wavelength element; the lack of flexure and fringing in DEIMOS allows one to
achieve Poisson-limited sky subtraction.  Finally, the spectra are smoothed
by a 10~pixel boxcar in wavelength weighted by their inverse variance.  Line
strength [equivalent width (EW)] measurements are performed on the normalized
spectra using a measurement window centered on the absorption feature
(e.g.,~\nai\ doublet at 8190~\AA) with continuum windows on each side of the
feature (see Table~\ref{tab:indexdef}).  Details of the EW measurements
are provided in the individual diagnostic sections (e.g.,~\S\,\ref{sec:na}).
The measured stellar radial velocities are corrected to heliocentric
velocities using the IRAF\footnote{IRAF is distributed by the National
Optical Astronomy Observatory, which is operated by the Association of
Universities for Research in Astronomy, Inc., under cooperative agreement
with the National Science Foundation.} task {\smcap rvcor}.
   
%%%%%%%%%%%%%%%%%%%%%%%%%%%%%%%%%%%%%%%%%%%%%%%%%%%%%%%%%%%%%%%%%%%%%%%%%%%%%
\subsection{Empirical Error Estimates from Duplicate
Measurements}\label{sec:error}
%%%%%%%%%%%%%%%%%%%%%%%%%%%%%%%%%%%%%%%%%%%%%%%%%%%%%%%%%%%%%%%%%%%%%%%%%%%%%

In some of our survey fields, masks were designed to overlap, allowing
for repeat observations of a total of 84~objects.  The duplicate measurements
allow us to make empirical estimates of the errors in the velocity and line 
widths measured from the spectra. 
Most of the 84~duplicate
measurements with successful redshifts are from field H13d (61).  There are
9~duplicate measurements with successful redshifts in field~a0, 8 in
field~a3, and 2 each in fields~H11 and~H13s.  For each pair of spectra, the
difference in the EW estimates (or radial velocity estimates) between
the two~masks is calculated.  The measurement error is assumed to be
proportional to the inverse of the S/N of the spectrum:
\begin{equation}
\sigma_{i}=C\left(\frac{N}{S}\right)_{i}.
\end{equation}
We define 
\begin{equation}
X_{ij}=\frac{(EW)_i-(EW)_j}{C\sqrt{\left(\frac{N}{S}\right)^{2}_{i}+\left(\frac{N}{S}\right)^{2}_{j}}}
\end{equation}
so that for the correct value of $C$, the distribution of $X_{ij}$'s will be
a Gaussian of unit width.  For each quantity measured from the spectra, 
Table~\ref{tab:errors} lists the error in that quantity for a S/N of 
10, which is typical of the spectra in our sample (see Table~\ref{tab:indexdef}
 for index definitions).  

%%%%%%%%%%%%%%%%%%%%%%%%%%%%%%%%%%%%%%%%%%%%%%%%%%%%%%%%%%%%%%%%%%%%%%%%%%%%%
\begin{table}[ht!]
\centering{
\caption{Empirical error estimates of quantities measured from the spectra,
$C\left(N/S\right)$, for our typical S/N of 10.}
\label{tab:errors}
\vspace{4mm}
\begin{tabular}{@{}lc@{}}
\hline\hline
Quantity        &  Typical Error   \\
\hline
Velocity  &  17~\kms\ \\
Na\,{\smcap i} EW & 0.74~\AA \\
Ca\,{\smcap ii} EW ($\rm\Sigma Ca$)  & 1.09~\AA  \\
$\rm[Fe/H]_{spec}$ & 0.46~dex  \\
K$_{7665}$ EW  & 1.4~\AA \\
K$_{7699}$ EW  & 0.6~\AA \\
TiO$_{7100}$ EW   & 0.07~\AA \\
TiO$_{7600}$(1) EW  & 0.05~\AA\\
TiO$_{7600}$(2) EW   & 0.1~\AA\\
TiO$_{7600}$(3) EW  &  0.3~\AA \\
TiO$_{7600}$(comb) EW  &  0.2~\AA \\
TiO$_{8500}$ EW  & 0.1~\AA \\
\hline\hline
\end{tabular}
}
\end{table}
%%%%%%%%%%%%%%%%%%%%%%%%%%%%%%%%%%%%%%%%%%%%%%%%%%%%%%%%%%%%%%%%%%%%%%%%%%%%%

In addition to providing empirical error estimates for line width and radial 
velocity measurements, the duplicate spectra were co-added to achieve higher 
S/N for measurement of the spectral features.  During coaddition, the 
individual spectra were 
weighted according to the measured inverse variance for each pixel.

%%%%%%%%%%%%%%%%%%%%%%%%%%%%%%%%%%%%%%%%%%%%%%%%%%%%%%%%%%%%%%%%%%%%%%%%%%
\section{Method for Selecting a Clean Sample of M31 Red
Giants}\label{sec:method}
%%%%%%%%%%%%%%%%%%%%%%%%%%%%%%%%%%%%%%%%%%%%%%%%%%%%%%%%%%%%%%%%%%%%%%%%%%%%%
To isolate a clean sample of RGB stars, we calculate empirical probability
distribution functions (PDFs) based on the properties of known RGB and dwarf
stars.  Each star is assigned individual probabilities of being an M31 RGB or
a MW dwarf star based on its location within each of the diagnostics.  The
probabilities from the first five diagnostics are then combined to give an
overall likelihood that the star
is an RGB star in M31 or a foreground dwarf star contaminant.  The selection
of RGB and dwarf star training sets, the diagnostics, construction of the
PDFs, and the final likelihood-based method are described below. 

%%%%%%%%%%%%%%%%%%%%%%%%%%%%%%%%%%%%%%%%%%%%%%%%%%%%%%%%%%%%%%%%%%%%%%%%%%%%%
\subsection{Training Set Selection}\label{sec:training_sets}
%%%%%%%%%%%%%%%%%%%%%%%%%%%%%%%%%%%%%%%%%%%%%%%%%%%%%%%%%%%%%%%%%%%%%%%%%%%%%
The \ddo-based preselection procedure we have used to select spectroscopic
targets is very effective at eliminating foreground MW dwarf stars in the
color range $1\lesssim(V-I)_0\lesssim2$, but is less effective outside this
color range (see \S\,\ref{sec:evidence} and \S\,\ref{sec:bluest}).  This
non-uniformity in the dwarf rejection probability as a function of color
would be difficult to characterize in a model training set.  We have
therefore chosen to use {\it empirical\/} RGB and dwarf training sets.  Since
our PDFs for each diagnostic are empirically determined, the selection of
unbiased and largely uncontaminated training sets is important.  Initial
training set stars were drawn from all fields observed during the 2002--2004
observing runs except field H13d, due to the difficulty in separating out M31
disk stars from Galactic dwarfs on the northeastern major axis (disk field).

The initial RGB training set stars were picked mainly on the basis of radial
velocity: $v<-300$~\kms\ (the systemic velocity of M31).  While it is
possible for MW halo stars to have a heliocentric radial velocity this
negative, the expected fraction in our overall data set is very small.  The
inital dwarf training set had to be chosen more carefully, since the radial
velocity distribution of M31 spheroid stars significantly overlaps the dwarf
velocity distribution. We included all stars which had been commented as
having strong \nai\ absorption during the data reduction phase.   Stars with
radial velocities $v>-200$~\kms\ in fields with a clear RGB/dwarf bimodality
in the radial velocity distribution were also included in the dwarf training
sets.  Finally, we included all the stars in field~m11, since its large
distance from M31 ($R\sim165$~kpc in projection) makes it extremely unlikely
for it to contain RGB stars based on previous estimates of the surface
brightness profile of M31's halo \citep{pri94}.

The initial training sets were used to calculate preliminary PDFs and
probabilities for all stars from the 2002--2004 observing runs.  We used
these preliminary RGB and dwarf probability estimates to isolate stars that
have a much higher combined probability (summed over the first five
diagnostics) of being an RGB star than a dwarf star.  These stars comprise
our final RGB star training set.  A similar procedure was used to construct
our final dwarf star training set.  These final RGB and dwarf star training
sets, consisting of a few hundred stars each, were used to refine the PDFs
for each diagnostic.

Given that a simple velocity cut alone is not enough to isolate a clean
sample of M31 RGB stars in the outer halo, it is worth explaining why we use
a velocity cut to select stars for the (initial) RGB training set:
\begin{itemize}
\item[$\bullet$]{The velocity cut used to define the initial RGB training
set, $v<-300$~\kms\ ($v<v_{\rm sys}^{\rm M31}$), is very extreme.  In fact,
only half of all the M31 RGB stars in our sample are expected to survive this
cut.  This is acceptable for the RGB training set, which only needs to be a
representative set of M31 RGB stars, not necessarily a complete one.  However,
such a cut would adversely affect our ability to find (and analyze the
dynamics of) RGB stars in M31's sparse outer halo.}
\item[$\bullet$]{There is a small but finite probability ($\lesssim1\%$) of
encountering a MW dwarf whose radial velocity is $v<-300$~\kms.  In the
outermost fields (e.g.,~field m11), where our sample includes of order 100~MW
dwarfs, this can translate to a few dwarf stars satisfying the velocity cut.
This is of the same order as the number of M31 RGB stars found in these
fields (see \S\,\ref{sec:subclass}, \S\,\ref{sec:rvcut}, and
Table~\ref{table:masks2}).  Thus, even this extreme velocity cut would lead
to a large MW dwarf contamination rate if we were to use it to search for M31
RGB stars in the outer fields.  By contrast these few MW dwarfs comprise a
negligible fraction ($<\!\!<1\%$) of the RGB training set; most RGB training
set stars come from the inner fields where the surface density of M31 RGB
stars greatly exceeds that of MW dwarf stars.}
\end{itemize}

To summarize, the initial training sets were based on extreme cuts in a
single parameter: radial velocity, Na line strength, or sky position.  This
resulted in limited, slightly contaminated, and somewhat biased sets of RGB
and dwarf stars.  By contrast, the final training sets were based on cuts in
the combined probability ratio.  Since this probability ratio combines
information from various parameters, the final training sets are more
complete, less contaminated, and less biased than the initial training sets.

The plots for the first five diagnostics below
(Figs.~\ref{fig:vel_diag}--\ref{fig:cmd_diag}) show the initial training set
stars overlaid on the final PDFs.  The fact that our PDFs did not change
drastically when we switched from the initial to final training sets suggests
that our training sets are stable.  We investigate our MW dwarf star training
set for possible biases by exploring the relationships between the different
parameters in \S\,\ref{sec:corr_dwf}.  The different properties of M31 RGB
training set stars --- \nai\ EW, \ki\ EW, TiO EW, \vio\ color, \ddo\
parameter, CMD position, and \fehp\ versus \fehs\ --- have no detectable
dependence on radial velocity.

%%%%%%%%%%%%%%%%%%%%%%%%%%%%%%%%%%%%%%%%%%%%%%%%%%%%%%%%%%%%%%%%%%%%%%%%%%%%%
\subsection{Diagnostics}\label{sec:diag}
%%%%%%%%%%%%%%%%%%%%%%%%%%%%%%%%%%%%%%%%%%%%%%%%%%%%%%%%%%%%%%%%%%%%%%%%%%%%%
%%%%%%%%%%%%%%%%%%%%%%%%%%%%%%%%%%%%%%%%%%%%%%%%%%%%%%%%%%%%%%%%%%%%%%%%%%%%%
\subsubsection{Method for Constructing Analytic Probability Distribution
Functions}\label{sec:pdf}
%%%%%%%%%%%%%%%%%%%%%%%%%%%%%%%%%%%%%%%%%%%%%%%%%%%%%%%%%%%%%%%%%%%%%%%%%%%%%
For each diagnostic, analytic PDFs are calculated for each of the M31 RGB and
MW dwarf star training sets.  In this section we describe the general process
of computing analytic PDFs; subsequent sections describe the specific
functions used for each diagnostic.  For the one-dimensional distributions
(radial velocity and \ddo\ parameter) an analytic function was fit to the
training set's distribution and normalized.  The PDFs for the two-dimensional
distributions (\nai\ EW, \ki\ EW, and TiO EW vs. ($V-I$) color, CMD position,
and \fehp\ versus\ \fehs) are also determined using an analytic fitting
scheme.  The distribution of the data along both the $x$ and $y$ axes must be
included in the final PDF. In modeling the two-dimensional PDF, the $y$
distribution is treated as a Gaussian whose mean and rms depend on $x$.  For
each training set, we compute the mean and rms of the $y$ distribution for
several bins in $x$: $\langle y\rangle_i$ and $\sigma_i^y$ as a function of
$x_i$, where $x_i$ is the mean value in each bin.  Analytic functions are
then fit to the measured $\langle y\rangle_i$ and $\sigma_i^y$ values, which
are denoted as $\langle y\rangle_{\rm fit}(x)$ and $\sigma_{\rm fit}^y(x)$,
respectively.  This describes how the training set's $y$ values are
distributed as a function of $x$, but does not contain information about the
distribution of $x$ values of the training set.  We therefore also fit a
smooth analytic function $P(x)$ to the projection of the training set
distribution along the $x$ axis.  Finally, the analytic two-dimensional PDF
is defined as: 
\begin{equation}
{\rm PDF}(x,\,y) = C~P(x)~\exp\Big[-0.5~[y~-~\langle y\rangle_{\rm
fit}(x)]^2~/~[\sigma_{\rm fit}^y(x)]^2\Big]
\label{eqn:pdf}
\end{equation}
where $C$ is a normalization constant such that the PDF integrates to unity.

%%%%%%%%%%%%%%%%%%%%%%%%%%%%%%%%%%%%%%%%%%%%%%%%%%%%%%%%%%%%%%%%%%%%%%%%%%%%%
\subsubsection{Radial Velocities}\label{sec:vel}
\begin{figure}
\centerline{\epsfxsize=6in \epsfysize=6in
\epsfbox{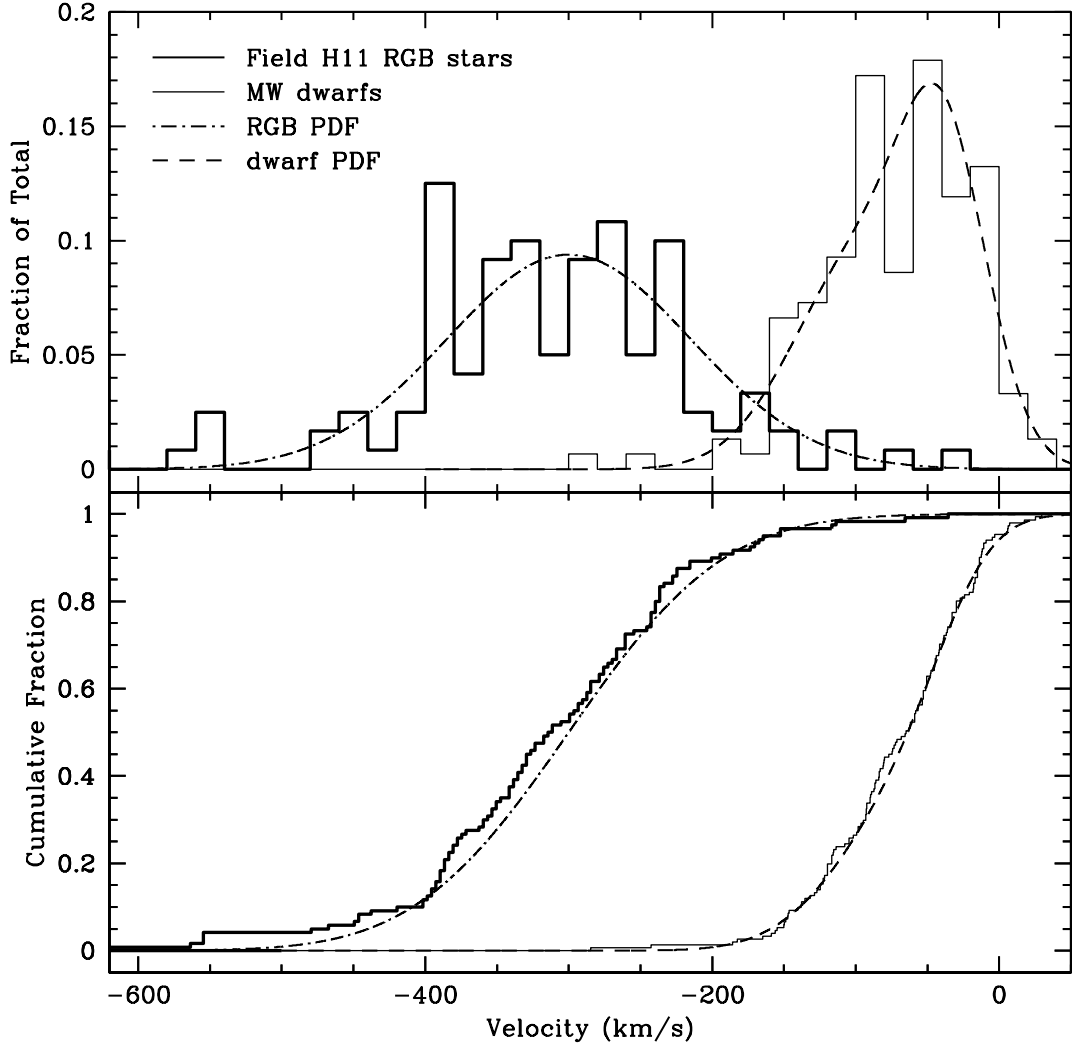}}
\figcaption[fig02.eps]{\label{fig:vel_diag}Probability distribution functions
for radial velocity in differential ({\it top\/}) and cumulative ({\it
bottom\/}) form.  The thin-lined histograms show the dwarf star training
set; the dashed curves show the double Gaussian fit used to define the PDF.
Since the red giant branch training set is selected on the basis of radial
velocity, it cannot be used to define the radial velocity PDF.  We instead
adopt a Gaussian as the RGB PDF (dot-dashed curve), centered on M31's
systemic velocity of $-300$~\kms\ with a width of $\sigma=85$~\kms, based on
a fit to the inner spheroid field~H11 sample (bold solid histograms).  Dwarf
stars lie close to $v_{\rm hel}\sim0$~\kms\ but have a negative tail due to
the reflex of the component of the Sun's velocity towards M31.  As such,
there is substantial overlap between the M31 RGB and MW dwarf radial velocity
distributions.
}
\end{figure}

\begin{figure}
\centerline{\epsfxsize=6in \epsfysize=6in
\epsfbox{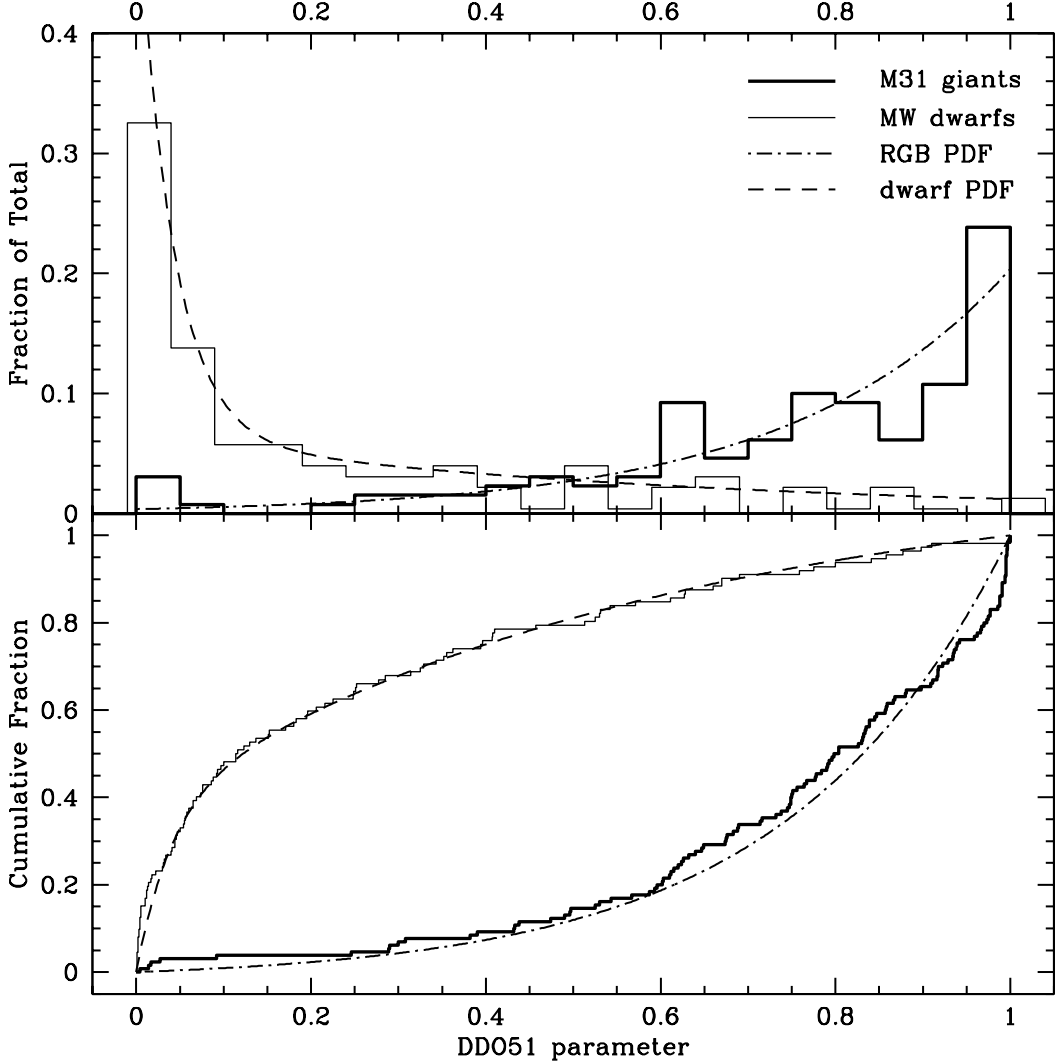}}
\figcaption[fig03.eps]{\label{fig:ddo_diag}Same as Figure~\ref{fig:vel_diag}
for the \ddo\ parameter.  The bold solid histograms represent the RGB
training set, a compilation of RGB stars from several fields close to M31's
center.  The \ddo\ parameter is assigned based on the position of a star in
the ($M-DDO51$) vs.\ ($M-T_2$) color-color diagram \citep{maj00}: a value
close to zero implies it is likely to be a dwarf star whereas a value close
to unity implies it is likely to be an RGB star.  The photometry is from KPNO
4-m Mosaic images in the Washington system $M$ and $T_2$ bands and in the
intermediate-width \ddo\ band, which includes the surface-gravity sensitive
Mg\,$b$ and MgH stellar absorption features at $\sim5100~$\AA.  Exponential
fits to the dwarf and RGB training sets are used to define the PDFs.  As
expected, the MW dwarf training set distribution peaks at zero and that of
the RGB training set peaks at unity, but there is significant overlap between
the two distributions.
}
\end{figure}

\begin{figure}
\centerline{\epsfxsize=5.5in \epsfysize=5.5in
\epsfbox{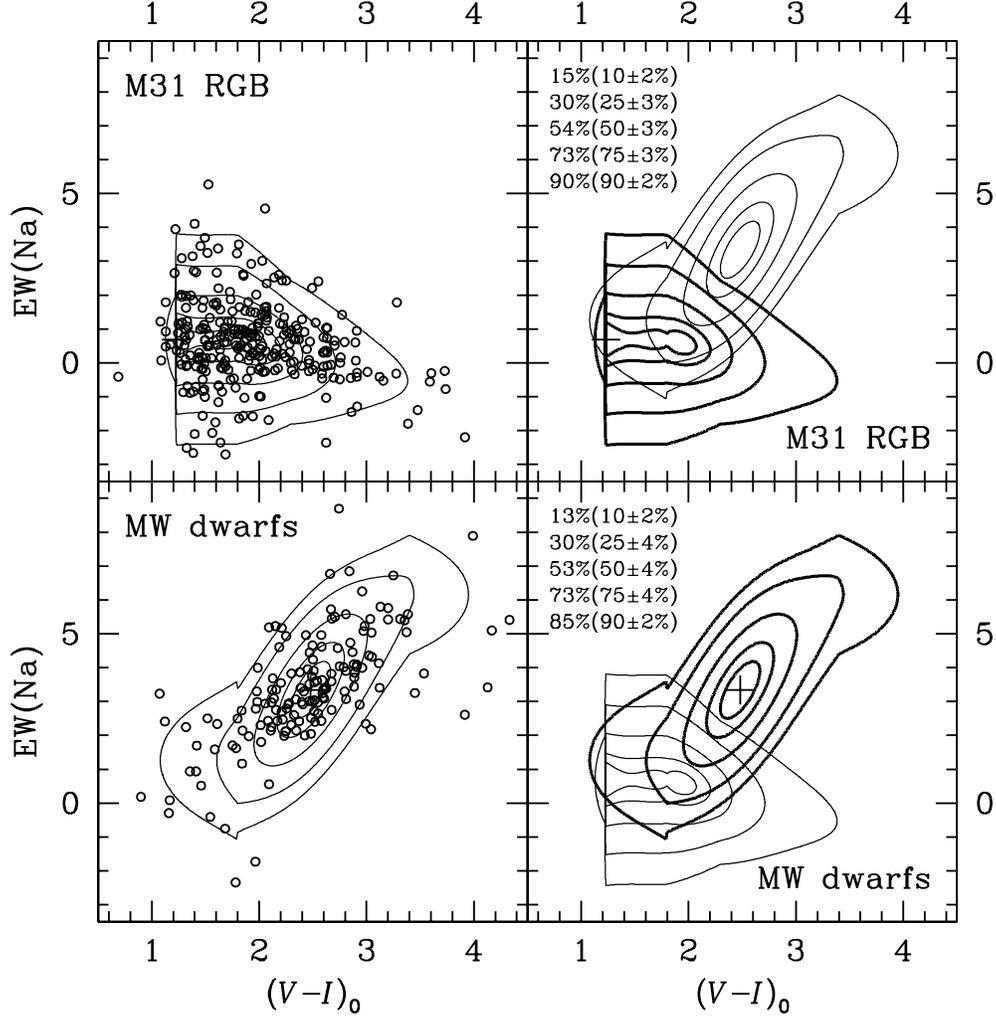}}
\figcaption[fig04.eps]{\label{fig:na_diag}Two-dimensional PDFs for the
equivalent width of the surface-gravity-sensitive \nai\ absorption line
doublet at 8190~\AA\ as a function of dereddened $(V-I)_0$ color (which
is sensitive to stellar effective temperature).  The left panels show
iso-probability contours for the RGB ({\it top\/}) and dwarf ({\it bottom\/})
PDFs, which are analytic fits to the distribution of training set stars (open
circles).  Both sets of PDF contours are shown in the right panels with the
RGB and dwarf contours in bold in the top and bottom panels, respectively.
The percentage of training set stars enclosed within each contour is
indicated in the right panels (ordered from innermost to outermost); the
percentage of analytic PDF space enclosed by the contour and the Poisson
error are listed within parentheses.  Cool dwarf stars with \vio\,$>2$
exhibit strong \nai\ absorption, while RGB and hotter dwarf stars show
negligible \nai\ absorption.  The color distribution of M31 RGB training set
stars is asymmetric; it peaks at \vio\,$\gtrsim1$ but has a tail extending
out to redder colors.  The MW dwarf training set stars have a somewhat
broader and symmetric color distribution centered on \vio\,$\approx2.5$.  The
\nai\ diagnostic is very effective at separating M31 RGB stars from MW dwarfs
for stars redder than \vio\,$\sim2.5$, but cannot distinguish between the two
stellar types for \vio\,$\lesssim2$.
}
\end{figure}

\begin{figure}
\centerline{\epsfxsize=5.5in \epsfysize=5.5in
\epsfbox{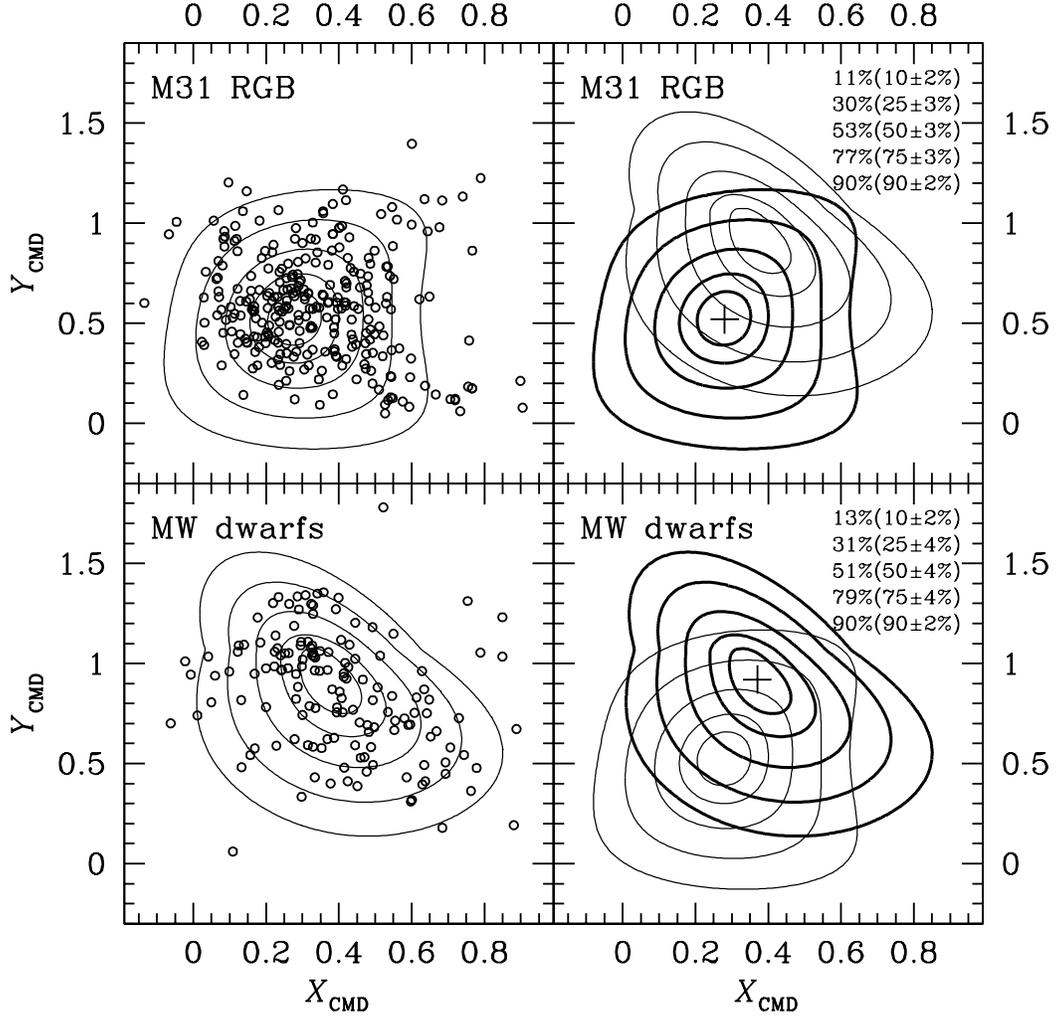}}
\figcaption[fig05.eps]{\label{fig:cmd_diag}Same as Figure~\ref{fig:na_diag}
for position within the \ivi\ color-magnitude diagram.  The abscissa is the
fractional distance across the isochrones: \xcmd\ = 0 at the most metal-poor
isochrone, $\rm[Fe/H]=-2.6$, and \xcmd\ = 1 at the most metal-rich isochrone,
$\rm[Fe/H]=+0.5$; the ordinate is the fractional distance along the RGB
isochrones: \ycmd\ = 0 at $I_0=22.5$, at the faint end of our spectroscopic
sample, and \ycmd\ = 1 at the tip of the RGB.  These parameters are
calculated using theoretical RGB isochrones from \citet{vdb06}, for an age of
$t=12$~Gyr and chemical composition of $\rm[\alpha/Fe]=0$, shifted to M31's
distance modulus of 24.47~mag.  Since M31 RGB stars tend to occupy the region 
of the CMD that is bounded by the isochrones (widest color spread near the
tip of the RGB, progressively smaller color spread toward fainter $I_0$
magnitudes), they form a roughly rectangular locus in (\xcmd, \ycmd) space.
On average, foreground dwarf stars tend to be brighter in $I_0$ than M31 RGB
stars.  Dwarfs are broadly distributed in the \ivi\ CMD, but since the RGB
isochrones curve toward redder colors as metallicity increases and tend to be
more widely separated toward the tip of the RGB, dwarf stars at higher \ycmd\
values tend to have a lower mean \xcmd\ and smaller spread in \xcmd.
}
\end{figure}

%%%%%%%%%%%%%%%%%%%%%%%%%%%%%%%%%%%%%%%%%%%%%%%%%%%%%%%%%%%%%%%%%%%%%%%%%%%%%
The radial velocity PDFs and histograms of the training set data are shown in
Figure~\ref{fig:vel_diag}.  Due to the \ddo\ pre-selection technique employed
in our survey (\S\,\ref{sec:slit_design}), dwarf distributions from standard
Galactic models \citep[e.g.,][]{bah84,rat85,rei02} cannot directly be used
for our data set.  The dwarf training set shows an asymmetric distribution,
and the sum of two Gaussians was fit to the heliocentric radial velocity
distribution of the dwarf training set (\S\,\ref{sec:dataredux}).  The
initial RGB training set was chosen on the basis of radial velocity
measurements, therefore it could not be used to define the RGB velocity PDF.
In addition, many of the observed fields have significant substructure or
small number statistics, making it difficult to choose an alternate training
set without biases.  Instead, we adopted a Gaussian centered on the systemic
velocity of M31 to define the RGB radial velocity PDF: $\langle{v}_{\rm
hel}\rangle=-300$~\kms\ and $\sigma=85$~\kms.  This is based on a fit to our
(admittedly limited) sample in H11, our smoothest spheroid field.
As shown in Figure~\ref{fig:vel_diag}, there is significant overlap between
the RGB and dwarf distributions, which makes it difficult to differentiate
between RGB and dwarf stars on the basis of radial velocity alone.  We will
return to this issue in \S\,\ref{sec:rvcut}.

Our adopted velocity dispersion of 85~\kms\ is probably not representative of
the entire M31 spheroid; previously published values are at least as large as
this \citep{rei02,p1,cha06}.  The velocity dispersion of M31's
outer halo ($R\gtrsim50$~kpc) remains poorly constrained at the present time.
It should be noted that the Gaussian sigma of 85~\kms\ is only used for the
definition of the RGB radial velocity PDF; we are {\it not\/} suggesting that
this be used as a model for the M31 bulge/halo.  Use of a larger sigma value
would result in a higher value for the RGB PDF in the overlap region between
MW dwarf and M31 RGB velocity distributions.  Thus, our adopted dispersion of
85~\kms\ represents a conservative choice.

In fields where a skewed distribution of M31 RGB velocities results in greater
overlap with the MW dwarf velocity distribution, the RGB velocity PDF can be
modified to account for this.  For example, the stellar disk of M31 is
apparent as a dynamically cold peak in the radial velocity histogram of our
NE major-axis field H13d.  We use the sum of two Gaussians to model the M31
disk and spheroid populations in that field.  The rest of our fields show no
evidence of a contribution from the extended stellar disk discussed by
\citet{iba05}.  This is not surprising given that most of our fields are
located at large distances from M31's center and/or lie close to its SE minor
axis \citep{kal06a}.

%%%%%%%%%%%%%%%%%%%%%%%%%%%%%%%%%%%%%%%%%%%%%%%%%%%%%%%%%%%%%%%%%%%%%%%%%%%%%
\subsubsection{\ddo\ Distribution}\label{sec:ddo}
%%%%%%%%%%%%%%%%%%%%%%%%%%%%%%%%%%%%%%%%%%%%%%%%%%%%%%%%%%%%%%%%%%%%%%%%%%%%%
The location of a star in ($M-DDO51$) versus ($M-T_2$) color-color space was
used to assign it a \ddo\ parameter
\citep[\S\,\ref{sec:slit_design};][]{pal03}.  Values of the \ddo\ parameter
close to 1 indicate that the star is probably an M31 RGB star while values
close to 0 indicate it is probably a foreground MW dwarf star.  The \ddo\
parameter was used to assign priorities to objects during the slitmask
design, and it can also be used to help differentiate between RGB and dwarf
stars independent of the spectroscopic data.  (In order to fill the masks to
the fullest extent possible, some objects with low \ddo\ parameters were
observed.)  Figure~\ref{fig:ddo_diag} shows the PDFs and the initial training
sets.  The dwarf training set was fit by a double exponential, and the RGB
training set was fit by a single exponential.  For both dwarf and RGB stars,
the training sets are reduced in number relative to the other diagnostics
because \ddo\ photometry was not available for fields~H11, H13s or H13d.  As
in the case of the radial velocity diagnostic, there is some overlap between
the RGB and dwarf PDFs.
 
%%%%%%%%%%%%%%%%%%%%%%%%%%%%%%%%%%%%%%%%%%%%%%%%%%%%%%%%%%%%%%%%%%%%%%%%%%%%%
\subsubsection{Na\,I Equivalent Width}\label{sec:na}
%%%%%%%%%%%%%%%%%%%%%%%%%%%%%%%%%%%%%%%%%%%%%%%%%%%%%%%%%%%%%%%%%%%%%%%%%%%%%

%%%%%%%%%%%%%%%%%%%%%%%%%%%%%%%%%%%%%%%%%%%%%%%%%%%%%%%%%%%%%%%%%%%%%%%%%%%%%
\begin{table}[ht!]
\centering{
\caption{Absorption line-strength index definitons used in this work.}
\label{tab:indexdef}
\vspace{4mm}
\begin{tabular}{@{}lccc@{}}
\hline\hline
Index          & Type\tablenotemark{a}&   Main        & Continuum        \\
               &     & Bandpass (\AA)  & Bandpasses (\AA) \\
\hline
Na\,{\smcap i} &$g~a$& 8179--8200 & 8130--8175, 8210--8220 \\
Ca\,{\smcap ii}$_{8498}$ &$g~a$ & 8489-8507 & 8560--8644  \\
Ca\,{\smcap ii}$_{8542}$ &$g~a$ & 8533-8551 &  8560--8644 \\
Ca\,{\smcap ii}$_{8662}$ &$g~a$ & 8653-8671 &  8560--8644  \\
K$_{7665}$     &$g~a$& 7660--7670 & 7677--7691, 7728--7741, 7764--7784, \\
               &     &            & 7802--7825, 7877--7895 \\
K$_{7699}$     &$g~a$& 7694--7703 & 7677--7691, 7728--7741, 7764--7784, \\
               &     &            & 7802--7825, 7877--7895 \\
TiO$_{7100}$   &$c~a$& 7055--7245 & 7012--7048, 7512--7576 \\
TiO$_{7600}$(1) &$c~a$& 7700--8000 & 7500--7580, 8080--8170 \\
TiO$_{7600}$(2) &$c~a$& 7736--7810 & 7500--7580, 8080--8170 \\
TiO$_{7600}$(3) &$g~s$&     none   & 7721--7748, 7802--7826, 7878--7895, \\
               &     &            & 7963--7986, 8001--8024, 8086--8113 \\
TiO$_{8500}$   &$g~s$&     none   & 8474--8484, 8563--8577, 8619--8642, \\
               &     &            & 8700--8725, 8776--8792 \\
\hline\hline
\end{tabular}
\tablenotetext{a}{Types $c~a$, $g~a$ and $g~s$ refer to classical-atomic,
generic-atomic, and generic-slope-like indices, respectively.}
}
\end{table}
%%%%%%%%%%%%%%%%%%%%%%%%%%%%%%%%%%%%%%%%%%%%%%%%%%%%%%%%%%%%%%%%%%%%%%%%%%%%%

The EW of the \nai\ absorption line at $\lambda=8190.5$~\AA\ is dependent on
both surface gravity and temperature, making it likely to be a useful
dwarf/RGB star discriminator \citep{sch97}.  Table \ref{tab:indexdef}
contains details of the EW measurement. 

This diagnostic relies on the distribution of stars in \vio\ color versus
\nai\ EW space.  Distributions in each quantity were fit analytically.  The
dwarf star training set's \vio\ color distribution was fit by a double
Gaussian and the RGB star training set's \vio\ color distribution was fit by
a Gaussian plus an exponential.  Each training set was then divided into bins
according to \vio\ color and the mean and rms of the \nai\ EW was calculated
for each bin.  Piecewise linear fits were made to the mean \nai\ EW vs.\
\vio\ and to the rms \nai\ EW vs.\ \vio.  The \nai\ EW distribution can be
characterized by Gaussians with running mean and sigma values based on the
linear fits.  The analytic fit to the \vio\ distribution was multiplied by
the (running) Gaussian fit to the \nai\ EW distribution to construct the
two-dimensional PDF.  In other words, $x$ and $y$ were set to \vio\ color and
\nai\ EW, respectively, in the context of \S\,\ref{sec:pdf} and
Eqn.~\ref{eqn:pdf}.

Figure~\ref{fig:na_diag} shows the iso-probability contours for the dwarf and
RGB PDFs, along with the position of the training set stars overlaid on the
contours.  Although there is some overlap in this diagnostic as well, there
is clear differentiation between dwarf and RGB stars at \vio\ colors redder
than $\sim2$, with dwarf stars showing increasingly strong \nai\ absorption. 
 
%%%%%%%%%%%%%%%%%%%%%%%%%%%%%%%%%%%%%%%%%%%%%%%%%%%%%%%%%%%%%%%%%%%%%%%%%%%%%
\subsubsection{CMD Position}\label{sec:cmd}
%%%%%%%%%%%%%%%%%%%%%%%%%%%%%%%%%%%%%%%%%%%%%%%%%%%%%%%%%%%%%%%%%%%%%%%%%%%%%
Position in the \ivi\ CMD provides yet another dwarf/RGB discriminator.  The
locii of RGB stars, spanning a wide range of metallicities and ages but all
at the same distance from us, form a well-defined shape in the CMD as
delineated by fiducial model RGB tracks (Fig.~\ref{fig:cmd}).  By contrast,
dwarf stars form a broad swath in color-magnitude space as they tend to be
spread out over a wide range of line-of-sight distances.  They also tend to
have brighter $I$ magnitudes.  To pursue this as a diagnostic, we defined two
parameters to describe a star's position in the CMD: \ycmd\ is the distance
of a star along an isochrone, running from 0 at $I_0=22.5$ (limiting
magnitude of our spectroscopic sample) to 1 at the tip of the RGB, and \xcmd\
is the distance of a star on a line drawn across the set of isochrones,
running from 0 at the most metal-poor isochrone used in our analysis,
$\rm[Fe/H]=-2.3$, to 1 at the most metal-rich isochrone $\rm[Fe/H]=+0.5$.
The lines across the isochrones are drawn by connecting points of roughly
equal mass bins.  Values of \ycmd\ $<$ 0 denote stars fainter than
$I_0\sim22.5$, and values of \ycmd\ $>$ 1 denote stars above the tip of the
RGB (values for these stars are calculated by linear extrapolation of the
isochrones).  Likewise, values of \xcmd\ $<$ 0 denote objects bluer than the
most metal-poor isochrone, and values of \xcmd\ $>$ 1 denote objects redder
than the most metal-rich isochrone (calculated by non-linear extrapolation of
the lines drawn through the isochrones).  The grid of isochrones used in this
analysis are from \citet{vdb06} and correspond to an age of $t=12.6$~Gyr and
$\rm[\alpha/Fe]=0$.

The distributions in \xcmd\ and \ycmd\ are combined to make the final
two-dimensional PDF.  In the context of \S\,\ref{sec:pdf} and
Eqn.~\ref{eqn:pdf}, setting $x$ to \ycmd\ and $y$ to \xcmd\ (instead of vice
versa) lends itself more naturally to our PDF construction scheme.
Specifically, single Gaussians were fit to the \ycmd\ distributions for the
RGB and dwarf training sets.  The stars were then divided into bins according
to their \ycmd values, and running Gaussian parameters were fit to the \xcmd\
distributions.  For each of the RGB and dwarf star training sets, the fits to
the \ycmd\ and \xcmd\ distributions were multiplied to make the final
two-dimensional PDF.  Figure~\ref{fig:cmd_diag} shows the iso-probability
contours for the PDFs and the dwarf and RGB training sets.  There is a lot of
overlap in this diagnostic, but there is still some differentiation between
the two populations.  Since the RGB stars follow the isochrones, they form a
rectangular distribution in \xcmd\ and \ycmd\ space.  In contrast, the dwarfs
are evenly distributed in the \ivi\ CMD, and since the metal-rich isochrones
curve to redder colors, the \xcmd\ values of bright dwarf stars become
progressively smaller with increasing \ycmd\ values.   

We have chosen to frame the CMD diagnostic in terms of (\xcmd,~\ycmd) instead
of the purely observational parameters $(I,~V-I)$.  While the values of
\xcmd\ and \ycmd\ are model dependent, the {\it separation\/} between the RGB
and dwarf PDFs is independent of the details of the theoretical RGB tracks
used to compute \xcmd\ and \ycmd.  For example, changing the adopted age or
$\rm[\alpha/Fe]$ for the model isochrones would introduce a small change in
the \xcmd\ (and \fehp) values but the change would be the same for RGB and
dwarf stars at any given location in the CMD.  Neither the separation between
the RGB and dwarf PDFs nor their widths would be affected by such a change.
The effectiveness of any diagnostic in isolating M31 RGB stars from MW dwarf
stars depends on the separation between the two PDFs relative to their widths.
The distribution of M31 RGB stars in (\xcmd,~\ycmd) space is
more physically meaningful than their CMD distribution: for example, there is
a one-to-one mapping from \xcmd\ to \fehp.  Moreover, \xcmd\ and \ycmd\ are
independent of each other in the RGB PDF and this makes it particularly easy
to construct the two-dimensional PDF within this space.

%%%%%%%%%%%%%%%%%%%%%%%%%%%%%%%%%%%%%%%%%%%%%%%%%%%%%%%%%%%%%%%%%%%%%%%%%%%%%
\subsubsection{Photometric versus Spectroscopic Metallicity
Estimates}\label{sec:feh}
%%%%%%%%%%%%%%%%%%%%%%%%%%%%%%%%%%%%%%%%%%%%%%%%%%%%%%%%%%%%%%%%%%%%%%%%%%%%%
The photometric metallicities ([Fe/H] estimates) for our data set are based
on the comparison of a star's position in the \ivi\ CMD with the above
grid ($\Delta Z = 0.0001$) of RGB isochrones at the distance of M31
\citep{vdb06}.  Since the photometric metallicity estimates are based on the
assumption that the stars are red giants at the distance of M31, it is
inherently incorrect for dwarf stars.  Hence, we expect a clear separation
between the photometric and spectroscopic [Fe/H] estimates for giants and
dwarfs.

Spectroscopic estimates for [Fe/H] are based on measurement of the \caii\
triplet at $\lambda\sim8500$~\AA.  The EW of the \caii\ triplet
($\rm\Sigma{Ca}$) was calculated using a linear combination of the EWs of the
three~lines that has been shown to maximize the S/N of the feature
\citep*{rut97a}:
\begin{equation}
\rm\Sigma{Ca}~~=~~0.5\,EW(\lambda8498~\AA)~+~1.0\,EW(\lambda8542~\AA)~+~
0.6\,EW(\lambda8662~\AA).
\end{equation}
The EW of the individual \caii\ lines was measured in
$\rm18~\AA$ wide bins.  \fehs\ was calculated from $\Sigma \rm Ca$ using an
empirical calibration relation:
\citep*{rut97b}:
\begin{equation}
{\rm[Fe/H]_{spec}} = -2.66 + 0.42\,[\Sigma{\rm Ca}-0.64(V_{\rm HB}-V)].
\end{equation}
The last term in the equation, $(V_{\rm HB}-V)$, corrects for the effect of
surface gravity based on the difference between the $V$ band apparent
magnitude and the apparent magnitude of M31's horizontal branch ($V_{\rm
HB}=25.17$) \citep{hol96}.  This calibration relation is derived from RGB
stars in MW globular clusters spanning a range of metallicities.
 
In the construction of the PDFs, $x$ and $y$ were set to \fehp\ and \fehs,
respectively (\S\,\ref{sec:pdf}).  Analytical fits were made to the dwarf and
RGB training set \fehp\ distributions.  Each training set was then divided
into \fehp\ bins, and Gaussians were fit to the \fehs\ distribution in each
bin.  A quadratic function was fit to the Gaussian parameters of each of the
RGB and dwarf star training sets.  The \fehs\ distribution was constructed
based on the running Gaussian parameters given by the quadratic fit.  For
each training set, the analytic fits to the \fehp\ and \fehs\ distributions
were multiplied to make the final two-dimensional PDF.
Figure~\ref{fig:cmd_diag} shows the PDF contours and training set
distributions in \fehs\ vs.\ \fehp.  As expected, the M31 RGB stars lie close
to the one-to-one relation while the dwarf stars are well removed from it;
there is a clear difference between the RGB and dwarf populations.

In using the above calibration relation, we are implicitly assuming that RGB
stars in M31 are similar to those in MW globular clusters in terms of having
old ages and $\rm[alpha/Fe]\sim+0.3$.  The latter is different from our value
of $\rm[\alpha/Fe]=0$ adopted in deriving the \fehp\ estimates for this
diagnostic.  In their study of M31's extended bulge, \citet{bro03} find that
in addition to the majority old population there is a significant
intermediate-age (6--8~Gyr) population present.  Any difference between our
adopted parameters and the unknown age and $\alpha$ enhancement of M31's halo
RGB stars will cause our \fehp\ and \fehs\ estimates to be offset from their
true values.  However, as discussed in \S\,\ref{sec:cmd}, such offsets will
not affect the effectiveness of the diagnostic.  For a more detailed treatment
of the M31 RGB metallicity estimates, we refer the reader to \citet{kal06a}.

%%%%%%%%%%%%%%%%%%%%%%%%%%%%%%%%%%%%%%%%%%%%%%%%%%%%%%%%%%%%%%%%%%%%%%%%%%%%%
\subsubsection{Other Diagnostics}\label{sec:othr_diag}
%%%%%%%%%%%%%%%%%%%%%%%%%%%%%%%%%%%%%%%%%%%%%%%%%%%%%%%%%%%%%%%%%%%%%%%%%%%%%

In this section we discuss two diagnostics based on \ki\ and three
diagnostics based on TiO.  These five diagnostics have a lot in common with
the \nai\ diagnostic and with one another (\S\,\ref{sec:na_k_tio}).  The
three TiO diagnostics have a smaller separation between RGB and dwarf stars
than the \nai\ or \ki\ diagnostics.  Our present method is set up for
combining diagnostics that are independent of one another so we have decided
to exclude the \ki\ and TiO diagnostics from the overall likelihood scheme
(\S\,\ref{sec:lcalc}).  We include them in this section because planned
improvements to our diagnostic method will enable us to incorporate them in
the future (\S\,\ref{sec:ind_diag}).

\noindent
{\bf K\,I Equivalent Width---}

The EWs of the \ki\ absorption lines at 7665 and $7699~$\AA, like the EW of
\nai, are dependent on temperature as well as surface gravity, making them
potential dwarf/giant discriminators.  The continuum bandpasses and main
bandpasses used to measure the EWs of the \ki\ absorption lines are listed in
Table~\ref{tab:indexdef}.  The construction of the two-dimensional RGB and
dwarf PDFs for each of the two \ki\ lines followed exactly the same procedure
as for the \nai\ line (\S\,\ref{sec:na}).

The iso-probability contours and the RGB and dwarf training sets are shown 
in Figure~\ref{fig:k_diag}. As in the \nai\ diagnostic, the M31 RGB and MW 
dwarf distributions overlap at bluer colors, but diverge at redder colors 
(\vio\ $> 2.5$), with the strength of the \ki\ absorption lines increasing 
with increasing \vio\ for dwarf stars.  

\bigskip
\bigskip
\noindent
{\bf Strength of TiO Bands---}

Absorption molecular bands are the strongest spectral features of cool stars.
In particular, TiO bands are apparent in early M-type stars and their
strength increases with decreasing temperature.  For a given temperature, TiO
bands of RGB stars are stronger than those of dwarf stars, making these
spectral features useful dwarf/giant discriminators in the low temperature
regime.  A detailed explanation of the behaviour of TiO bands (e.g.,~those
around $\lambda\sim8500$ and 8900~\AA) is provided in \citet{cen01,cen06} on
the basis of an extensive, empirical stellar library in the near-infrared.

We have measured the TiO bands at $\lambda \sim 7100$, 7600, and 8500~\AA\
using the different index definitions listed in Table~\ref{tab:indexdef}.
TiO$_{7100}$ is a classic atomic index for the TiO band at
$\lambda\sim7100$~\AA\ that consists of two continuum bandpasses and one
central bandpass.  For the TiO band at $\lambda \sim 8500$~\AA\ we have
followed the {\sl generic}\footnote{{\sl Generic} indices, defined as those
having multiple continuum and central bandpasses, are specially suited to
measure spectral features in regions crowded by other contaminating features,
sky emission line residuals, and telluric absorption.} slope-like index
definition of \citet{cen03}.  In short, the signal of pixels within five
continuum bandpasses (see Table~\ref{tab:indexdef}) are used to fit a
straight line---a pseudo-continuum, $F_C(\lambda)$---over the region
dominated by the TiO bands.  The TiO$_{8500}$ index is thus computed as the
ratio between the pseudo-continuum values at the central wavelengths of the
reddest and bluest continuum bandpasses:
TiO$_{8500}=F_C(\lambda8784.0)/F_C(\lambda8479.0)$.  In a sense, it may be
considered to be a measure of the local pseudo-continuum slope, so
TiO$_{8500}$ values around 1 mean that no TiO band is present in the spectra.
Finally, the strength of the TiO band at $\lambda \sim 7600$~\AA,
TiO$_{7600}$, is defined to be the mean of three normalized, different index
definitions: TiO$_{7600}$(1), TiO$_{7600}$(2), and TiO$_{7600}$(3).
TiO$_{7600}$(1) and TiO$_{7600}$(2) are classic atomic index
definitions---the latter having a narrower central bandpass---whereas
TiO$_{7600}$(3) [$=F_C(\lambda7734.5)/F_C(\lambda8099.5)$] is another
slope-like index defined by means of six continuum bandpasses (see
Table~\ref{tab:indexdef}).  Since the spectra of our target stars are usually
affected by inter-CCD gaps located around this spectral region, we preferred
to employ different index definitions in order to ensure a large number of
training set stars having available index measurements for this TiO band.
Thus, in case any of the above index definitions was not reliable or
impossible to measure, means of the remaining reliable measurements were
computed to derive the final TiO$_{7600}$ value.

Two-dimensional RGB and dwarf PDFs for each of the three TiO diagnostics were
constructed following the same procedure as for the \nai\ line
(\S\,\ref{sec:na}).  Figure~\ref{fig:tio_diag} shows the iso-probability
contours for the dwarf and RGB PDFs, along with the position of the training
set stars overlaid on the contours.  It is clear that TiO diagnostics are not
really useful for \vio\ colors $\la 2.0$ (since no apparent TiO bands are
found for those temperatures).  However, the TiO strengths of dwarf and RGB
stars follow two different increasing trends with increasing \vio.  Since
TiO$_{7600}$ might be affected by nearby telluric absorption, the
discriminating power of this index is not as clear cut as it is for
TiO$_{7100}$ and TiO$_{8500}$.

%%%%%%%%%%%%%%%%%%%%%%%%%%%%%%%%%%%%%%%%%%%%%%%%%%%%%%%%%%%%%%%%%%%%%%%%%%%%%
\subsection{Computation of Likelihood Values}\label{sec:lcalc}
%%%%%%%%%%%%%%%%%%%%%%%%%%%%%%%%%%%%%%%%%%%%%%%%%%%%%%%%%%%%%%%%%%%%%%%%%%%%%
For each star in our survey with a measured blueshift, the probability the
star is a red giant (\pg) or a dwarf (\pd) is calculated based on the 
normalized RGB and dwarf star PDFs.  The likelihood a star $i$ is a 
red giant in a given diagnostic $j$ is computed using the formula
\begin{equation}\label{eqn:lij}
L_{ij}~=~\log\left(\frac{P_{\rm giant}}{P_{\rm dwarf}}\right).
\end{equation}
%where $R$ is the expected ratio between the number of RGB and dwarf stars
%in a given field.  For each field, the
%shape of the RGB and dwarf PDFs remains the same, but the relative number of
%expected RGB stars to dwarfs changes, since the number of dwarfs should remain
%relatively constant while the RGB surface denisty decreases with increasing
%distance from M31.  Including the factor $R$ in the computation of $L_i$ has
%the effect of scaling the normalized RGB and dwarf PDFs relative to each other.  
To determine a star's overall likelihood value, we compute a weighted average
of all the individual likelihoods available for each star:
\begin{equation}
\langle{\it L}_{\it i}\rangle= \frac{\displaystyle\sum_j w_j L_{ij}}{\displaystyle\sum_j w_j},
\end{equation}
summed over the available diagnostics.  In general, all available individual
likelihoods for a given star receive equal weight, $w_j=1$.  The only
exception to this is for stars that are outliers in {\it both\/} the RGB and
dwarf PDFs in any of two-dimensional diagnostics (\nai\ EW, CMD position, and
\fehp\ vs. \fehs).  For such stars the corresponding diagnostic is assigned a
lower weight ($w_j<1$) as described below.  The purpose of this is to
downweight, and in extreme cases effectively remove from the overall
likelihood estimate, diagnostics in which the star lies in a region of
parameter space that is poorly sampled by both the RGB and dwarf star training
sets.  For the same reason, the individual likelihoods are capped at $\pm5$,
which corresponds to a \pg\ to \pd\ ratio of $10^5$ or $10^{-5}$.

Specifically, this downweighting occurs for stars that satisfy the criteria:
$P_{\rm giant}<3\sigma_{\rm giant}$ and $P_{\rm dwarf}<3\sigma_{\rm dwarf}$ in
a given diagnostic, where the probability thresholds $\sigma_{\rm giant}$ and
$\sigma_{\rm dwarf}$ represent the probability levels which include 90\% of
the RGB and dwarf training set stars, respectively.  The weight $w_j$ assigned
to such a star in the $j$-th diagnostic is:
\begin{equation}\label{eqn:weight}
w_j = \frac{w_0}{\left(\frac{\sigma}{P}\right)_{\rm
giant}^2+\left(\frac{\sigma}{P}\right)_{\rm dwarf}^2},
\end{equation}
where $w_0$ is set to 2/9 to ensure that the weights are well-behaved for
stars that are close to the above thresholds.  We also investigated weighting
the individual likelihoods for all stars in all diagnostics based on the
measurement errors in the relevant parameters (\S\,\ref{sec:error}), but
found no significant differences in the overall likelihood distributions.  
 
We found that the addition of the \ki\ and TiO diagnostics did not improve
our ability to separate RGB and dwarf stars, and in fact hurt it by putting
too much weight on \vio\ colors.  There is strong covariance among the \nai,
\ki, and TiO diagnostics because they share a common color axis (see
\S\,\ref{sec:na_k_tio} for more details).  The three TiO diagnostics are not
as sensitive as the \nai\ diagnostic---the vertical separation between the
dwarf and RGB PDFs in the TiO diagnostics is relatively small
(Fig.~\ref{fig:tio_diag})---and the degeneracy between the RGB and dwarf PDFs
for hot stars in the \ki\ diagnostics extends to redder colors than in the
\nai\ diagnostic (compare Figs.~\ref{fig:na_diag} and \ref{fig:k_diag}).
Furthermore, the fractional measurement error in the EW is larger for \ki\
and TiO than for \nai: the \ki\ lines and TiO$_{7600}$ band are affected by
the atmospheric A-band, and the two reddest TiO bands are occasionally
affected by spectral continuum discontinuity artifacts at the inter-CCD
interface in DEIMOS.

For these reasons we do {\it not\/} include the \ki\ and TiO diagnostics in
our final calculation of the overall likelihood.  We have nevertheless
decided to present the details of the five \ki\ and TiO diagnostics as they
may be useful for other data sets.  Moreover, in \S\,\ref{sec:ind_diag} we
discuss future modifications that should render the \nai, \ki, and TiO
diagnostics independent of one another.

%%%%%%%%%%%%%%%%%%%%%%%%%%%%%%%%%%%%%%%%%%%%%%%%%%%%%%%%%%%%%%%%%%%%%%%%%%%%%
\subsection{Overall Likelihood Distributions}\label{sec:ldist}
%%%%%%%%%%%%%%%%%%%%%%%%%%%%%%%%%%%%%%%%%%%%%%%%%%%%%%%%%%%%%%%%%%%%%%%%%%%%%
Figure \ref{fig:lhist} shows the distribution of overall likelihood values
for each of our Keck/DEIMOS fields.  These histograms use all available
diagnostics (excluding \ki\ and TiO) for each star.  

The overall likelihood distributions for each field were used to determine
whether a star was a red giant or a dwarf.  The overall likelihood
($\langle{L_i}\rangle$) histograms in Figure~\ref{fig:lhist} show obvious
peaks at $\gtrsim+1$ (RGB) and $\lesssim+1$ (dwarf).  The relative height of
the peaks change with radius: in fields with approximately equal numbers of
RGB and dwarf stars we see a clear bimodality (ie. field a13), while for
inner fields we see a strong peak of RGB stars with a tail of dwarf stars,
and vice versa for the outer fields.  Stars with $\langle{L_i}\rangle>0$ are
designated M31 RGB stars, and stars with $\langle{L_i}\rangle<0$ are
designated foreground MW dwarf stars.
 
Figure~\ref{fig:prob} illustrates the power of our RGB/dwarf separation
technique using stars in the $R=30$~kpc minor-axis field~a0.  The individual
likelihoods---defined as $L_j\equiv(P_{\rm giant}/P_{\rm dwarf})$ for the
$j$-th diagnostic---for the five primary diagnostics are plotted against one
another ($L_j$ vs.\ $L_k$).  Histograms for each individual likelihood $L_j$
are plotted in the panels along the lower-right diagonal.  It is encouraging
to see that stars that are classified as M31 RGB stars by our
likelihood-based method (blue points) tend to lie in the top right quadrant
of all panels ($L_j,\,L_k>0$) while those classified as MW dwarf stars (red
points) tend to lie in the bottom left quadrant ($L_j,\,L_k<0$).  No single
diagnostic is able to discriminate perfectly between M31 RGB and MW dwarf
stars, however the combination of diagnostics is very effective at separating
the two stellar types.

Although there is no explicit weighting of the diagnostics 
(\S\,\ref{sec:lcalc}), some are more powerful than others.  Diagnostics with a
large separation between the RGB and dwarf PDFs (relative to the widths of the
PDFs) will have a large range of individual likelihood values and vice versa
(e.g.,~compare the range of $L_v$ vs.\ $L_{\rm CMD}$ values in
Fig.~\ref{fig:prob}).  Thus, diagnostics with a small separation between the
dwarf and RGB PDFs will implicitly have less of an effect on the overall
likelihood value.

The overall range of $L_{\rm CMD}$ and $L_{\rm DDO51}$ is small with respect
to the other diagnostics.  This is because of the substantial overlap between
the RGB and dwarf PDFs for these two diagnostics.  Despite the small range,
there appears to be a fairly clear RGB/dwarf separation in both of these
diagnostics, indicating that our measurement error in these quantities is
small compared to the width of the corresponding distribution.  In field~a0
the $L_{\rm Fe/H}$ diagnostic appears to be the least discriminatory of all
the diagnostics.  This is because of instrumental/data reduction problems
with our early DEIMOS spectra (most of our a0 spectra are from fall 2004).
The \caii\ triplet in particular is severely affected by sky subtraction
errors in the a0 spectra.  By contrast the \nai\ doublet is in a part of the
spectrum free of strong night sky emission lines and is well removed from the
DEIMOS inter-CCD gap so the $L_{\rm Na}$ diagnostic is well behaved in
field~a0.  Despite its relative poor performance in field~a0, the $L_{\rm
Fe/H}$ diagnostic in general works better than the CMD diagnostic and as well
as the DDO51 diagnostic.

%%%%%%%%%%%%%%%%%%%%%%%%%%%%%%%%%%%%%%%%%%%%%%%%%%%%%%%%%%%%%%%%%%%%%%%%%%%%%
\begin{table*}[ht!]
\begin{center}
\caption{Comparing the effectiveness of the five primary
diagnostics.\tablenotemark{a}}
\vskip 0.3cm
\begin{tabular}{lrrrrrrrrrrrr}
\hline
\hline
\multicolumn{1}{l}{Diagnostic(s)} &
  \multicolumn{6}{c}{$R=30$--60~kpc} &
  \multicolumn{6}{c}{$R=60$--165~kpc} \\
& \multicolumn{3}{c}{Secure RGB (142)} &
  \multicolumn{3}{c}{Secure Dwarf (70)} &
  \multicolumn{3}{c}{Secure RGB (22)} &
  \multicolumn{3}{c}{Secure Dwarf (202)} \\
& \multicolumn{1}{c}{Corr.} &
  \multicolumn{1}{c}{Unc.} &
  \multicolumn{1}{c}{Inc.} &
  \multicolumn{1}{c}{Corr.} &
  \multicolumn{1}{c}{Unc.} &
  \multicolumn{1}{c}{Inc.} &
  \multicolumn{1}{c}{Corr.} &
  \multicolumn{1}{c}{Unc.} &
  \multicolumn{1}{c}{Inc.} &
  \multicolumn{1}{c}{Corr.} &
  \multicolumn{1}{c}{Unc.} &
  \multicolumn{1}{c}{Inc.} \\
\hline
$L_v$           &  96.5 &   2.8 &   0.7 &  85.7 &  11.4 &   2.9 &
                   90.9 &   4.5 &   4.5 &  92.6 &   5.0 &   2.5 \\
$L_{\rm DDO51}$ &  70.4 &  26.1 &   3.5 &  64.3 &  20.0 &  15.7 &
                   68.2 &  27.3 &   4.5 &  73.3 &  20.3 &   6.4 \\
$L_{\rm Na}$    &  80.3 &  16.9 &   2.8 &  68.6 &  25.7 &   5.7 &
                   77.3 &  22.7 &   0.0 &  71.3 &  21.3 &   7.4 \\
$L_{\rm CMD}$   &  44.4 &  54.9 &   0.7 &  21.4 &  72.9 &   5.7 &
                   63.6 &  36.4 &   0.0 &  24.8 &  73.3 &   2.0 \\
$L_{\rm Fe/H}$  &  66.2 &  23.9 &   9.9 &  85.7 &  11.4 &   2.9 &
                   54.5 &  27.3 &  18.2 &  79.2 &  17.8 &   3.0 \\
$\langle L\rangle_2$\tablenotemark{b}
                &  94.4 &   4.9 &   0.7 &  84.3 &  14.3 &   1.4 &
                   90.9 &   9.1 &   0.0 &  90.1 &   7.4 &   2.5 \\
$\langle L\rangle_3$\tablenotemark{c}
                &  95.8 &   4.2 &   0.0 &  94.3 &   4.3 &   1.4 &
                   90.9 &   9.1 &   0.0 &  95.5 &   3.0 &   1.5 \\
$\langle L\rangle_4$\tablenotemark{d}
                & 100.0 &   0.0 &   0.0 &  94.3 &   4.3 &   1.4 &
                  100.0 &   0.0 &   0.0 &  96.5 &   2.0 &   1.5 \\
\hline
\end{tabular}
\tablenotetext{a}{Percentage of secure RGB/dwarf stars in a given radial bin
   that were correctly classified by a given diagnostic/combination of
   diagnostics (``Corr.''), had an uncertain classification (``Unc.''), or
   were incorrectly classified (``Inc.''); see text for details.  Secure RGB
   stars are defined to be Class~+3 and +2 stars while secure dwarf stars are
   defined to be Class~$-$3 and $-$2 stars (see \S\,\ref{sec:subclass} and
   Table~\ref{table:masks2}); the total number in each radial bin is given
   in parentheses after the column headings.}
\tablenotetext{b}{Weighted average of the two best diagnostics: $L_v$ and
$L_{\rm Na}$.}
\tablenotetext{c}{Weighted average of the three best diagnostics: $L_v$, 
$L_{\rm Na}$, and $L_{\rm DDO51}$.}
\tablenotetext{d}{Weighted average of the four best diagnostics: $L_v$, 
$L_{\rm Na}$, $L_{\rm DDO51}$, and $L_{\rm Fe/H}$.}
\label{table:diags}
\end{center}
\end{table*}
%%%%%%%%%%%%%%%%%%%%%%%%%%%%%%%%%%%%%%%%%%%%%%%%%%%%%%%%%%%%%%%%%%%%%%%%%%%%%

We next attempt to quantify how each individual diagnostic performs on the
secure samples of M31 RGB stars and MW dwarf stars (Class~+3/+2 and
$-$3/$-$2, respectively; see \S\,\ref{sec:subclass}).  Depending on whether
the individual likelihood value for the $j$-th diagnostic and $i$-th secure
M31 RGB star [as defined by Eqn.~(\ref{eqn:lij})] is $>+0.5$, between $-0.5$
and +0.5, or $<-0.5$, that star is deemed to have a correct, uncertain, or
incorrect classification.  Similar classifications are attempted for secure
MW dwarf stars using individual diagnostics.  The percentages of correct,
uncertain, and incorrect classifications are listed in
Table~\ref{table:diags} for each of the five primary diagnostics using secure
RGB/dwarf stars in two broad radial bins: $R=30$--60~kpc (fields~a0, a3, and
a13) and $R=60$--165~kpc (fields~a19, m6, b15, m8, and m11).  The radial
velocity diagnostic is the most effective (highest percentage of correct
classifications; the lowest percentage of uncertain/incorrect classifications)
while the CMD diagnostic is the least effective.  The \nai\ EW diagnostic is
the second best; the DDO51 and \fehp\ vs.\ \fehs\ diagnostics are comparable.
Weighted averages of various subsets of diagnostics are also tested; their
percentages are listed next in Table~\ref{table:diags}.

%%%%%%%%%%%%%%%%%%%%%%%%%%%%%%%%%%%%%%%%%%%%%%%%%%%%%%%%%%%%%%%%%%%%%%%%%%%%%
\subsection{Subclassification of the Red Giant and Dwarf Star
Samples}\label{sec:subclass} 
%%%%%%%%%%%%%%%%%%%%%%%%%%%%%%%%%%%%%%%%%%%%%%%%%%%%%%%%%%%%%%%%%%%%%%%%%%%%%
In general, stars with $\langle{L_i}\rangle>0$ are M31 RGB stars while those
with $\langle{L_i}\rangle<0$ are foreground MW dwarf stars.
We have developed a classification scheme which divides each of the RGB and
dwarf samples into three classes, with Class~$+3$ ($-3$) being the most
secure and Class~$+1$ ($-1$) the least secure.  The classes are based
primarily on $\langle{L_i}\rangle$, but \xcmd\ is used as a secondary
criterion (\S\,~\ref{sec:cmd}).  The most secure RGB (Class~$+3$) and dwarf
stars (Class~$-3$) are identified as those with $\langle{L_i}\rangle>0.5$ and 
$\langle{L_i}\rangle<-0.5$, respectively.  However, to be classified as a
Class~$+3$ object, a star must also pass the \xcmd$\ge0.0$ criterion, which
ensures that it falls within the color range of theoretical RGB isochrones at
the distance of M31. 

Stars with $\langle{L_i}\rangle>0.5$ and $-0.05\le$\,\xcmd\,$<0$ are
designated Class $+2$.  The errors in our \xcmd\ measurement are
approximately 0.05 (the errors are a little larger than this at the faint end
of our sample and smaller at the bright end).  Thus, Class~$+2$ stars are
those with large $\langle{L_i}\rangle$ values that lie slightly bluer than
(but roughly consistent with) the most metal-poor isochrone.

%%%%%%%%%%%%%%%%%%%%%%%%%%%%%%%%%%%%%%%%%%%%%%%%%%%%%%%%%%%%%%%%%%%%%%%%%%%%%
\begin{table*}[ht!]
\begin{center}
\caption{Confirmed M31 RGB and MW dwarf stars for each of the
fields.\tablenotemark{a}}
\vskip 0.3cm
\begin{tabular}{lccccccccl}
\hline
\hline
\multicolumn{1}{c}{Field} & \multicolumn{1}{c}{$R$} & \multicolumn{1}{c}{No.
Sci.}  & \multicolumn{3}{c}{No. M31 RGB Stars} &
\multicolumn{3}{c}{No. MW Dwarf Stars} & \multicolumn{1}{c}{Comments} \\
& \multicolumn{1}{c}{(kpc)} & \multicolumn{1}{c}{Spectra\tablenotemark{b}} & \multicolumn{1}{c}{Class} & \multicolumn{1}{c}{Class} & \multicolumn{1}{c}{Class} & \multicolumn{1}{c}{Class} & \multicolumn{1}{c}{Class} & \multicolumn{1}{c}{Class} & \\
& & &\multicolumn{1}{c}{+3} & \multicolumn{1}{c}{+2} & \multicolumn{1}{c}{+1} &
\multicolumn{1}{c}{$-3$} & \multicolumn{1}{c}{$-2$} & \multicolumn{1}{c}{$-1$}  \\
\hline
H11    & 12  & 234 &  102 & 0 & 13 & 17 & 1 &  6 &\\
H13s   & 21  & 221 &   83 & 1 & 20 & 18 & 1 &  6 & giant southern stream field \\
H13d\tablenotemark{c}
       & 25  & 136 &   67 &...&... & 19 &...&... & disk field\\
a0     & 30  & 245 &   61 & 4 & 10 & 30 & 2 &  3 & \\
a3     & 33  & 224 &   59 & 0 &  8 & 13 & 0 &  4 & giant southern stream field \\
a13    & 60  & 141 &   16 & 2 &  2 & 23 & 2 &  3 & \\
a19    & 81  & 68  &    4 & 0 &  2 & 23 & 0 &  3 &\\
m6     & 87  & 138 &    7 & 2 &  0 & 44 & 3 &  8 &\\
b15    & 95  & 127 &    5 & 0 &  2 & 24 & 1 & 10 & \\
m8     & 121 & 109 &    1 & 0 &  2 & 22 & 2 &  5 & \\
m11    & 165 & 283 &    3 & 0 &  4 & 82 & 1 & 11 & \\

\hline
\end{tabular}
\tablenotetext{a}{The subclassification of the M31 RGB stars into Class +3,
   +2, and +1 and MW dwarf stars into Class $-$3, $-$2, and $-$1 (very
   secure, secure, and marginal) is discussed in \S\,\ref{sec:subclass}.}
\tablenotetext{b}{Number of unique science spectra, excluding duplicate
   measurements (\S\,\ref{sec:error}), $Q=-2$ instrumental failures
   (\S\,\ref{sec:dataredux}), and alignment stars.}
\tablenotetext{c}{The photometry in our H13d field is currently undergoing
   recalibration so we do not attempt to subclassify the M31 RGB stars and MW
   dwarf stars at this point.}
\label{table:masks2}
\end{center}
\end{table*}
%%%%%%%%%%%%%%%%%%%%%%%%%%%%%%%%%%%%%%%%%%%%%%%%%%%%%%%%%%%%%%%%%%%%%%%%%%%%%

The least secure/marginal M31 RGB stars (Class~$+1$) are defined to be those
with $0.0<\langle{L_i}\rangle\le0.5$ and \xcmd\,$\ge-0.05$.  Likewise, the
marginal MW dwarf stars (Class~$-1$) are those with
$-0.5\ge\langle{L_i}\rangle\le0.0$ and \xcmd\,$\ge-0.05$.  We expect some
misclassification of stars within the range
$-0.5\le\langle{L_i}\rangle\le0.5$, because stars with $\langle{L_i}\rangle$
values in this range often have a spread of individual likelihood values,
some positive (they follow the RGB PDF) and some negative (they follow the
dwarf PDF).  

Finally, stars with $\langle{L_i}\rangle\ge-0.5$ and \xcmd\,$<-0.05$ are
designated Class~$-2$ objects.  These objects fall considerably blueward of
the most metal-poor theoretical isochrone in the CMD.  We disregard the
$\langle{L_i}\rangle$ values of these stars because the diagnostic method
does not work well for such blue objects
(\S\S\,~\ref{sec:bluest}--\ref{sec:marg}).  

This classification scheme allows us to systematically identify marginal
stars and remove them from our analysis.  Table~\ref{table:masks2} lists the
number of M31 RGB and MW dwarf stars per class in each field.
Figure~\ref{fig:cmd}~({\it top and middle\/}) shows the distribution in the
CMD of the three RGB classes and three dwarf classes, respectively.

%%%%%%%%%%%%%%%%%%%%%%%%%%%%%%%%%%%%%%%%%%%%%%%%%%%%%%%%%%%%%%%%%%%%%%%%%%%%%
\section{Discussion of the Method}\label{sec:disc} 
%%%%%%%%%%%%%%%%%%%%%%%%%%%%%%%%%%%%%%%%%%%%%%%%%%%%%%%%%%%%%%%%%%%%%%%%%%%%%
\subsection{Statistical Tests}\label{sec:tests}
%%%%%%%%%%%%%%%%%%%%%%%%%%%%%%%%%%%%%%%%%%%%%%%%%%%%%%%%%%%%%%%%%%%%%%%%%%%%%
\subsubsection{Evidence of M31 RGB Stars in the Outermost
Fields}\label{sec:evidence}
%%%%%%%%%%%%%%%%%%%%%%%%%%%%%%%%%%%%%%%%%%%%%%%%%%%%%%%%%%%%%%%%%%%%%%%%%%%%%

Is it possible that the stars identified as secure M31 RGB stars in our
outermost fields are actually misidentified MW dwarf stars (e.g.,~stars in
tails/outskirts of the dwarf star distributions in the various diagnostic
parameter spaces)?  We do not know {\it a priori\/} what fraction of the
overall dwarf population is in the form of such outliers, but expect the
fraction to be the same in all fields.  By contrast, the ratio of secure RGB
stars to secure MW dwarfs is observe to decline monotonically with increasing
projected distance from M31's center \citep{guh05}.

Given the very small number of secure M31 RGB stars found in our outermost
halo fields, it is very important to check whether the properties of these
stars are similar to those of the bulk of M31 RGB stars in our sample.
Figure~\ref{fig:outrmst_4pnl} compares the distribution of very secure M31 RGB 
(Class~+3) stars in our two outermost fields~m8 and m11 to that of foreground
MW dwarf stars in four of the diagnostic plots.  The handful of very secure
M31 RGB stars and the far more numerous secure M31 dwarf stars follow the RGB
and dwarf PDFs, respectively.  This suggests that the Class~$+3$ objects found
in these fields are indeed genuine M31 RGB stars.

Figure~\ref{fig:clr_dist} shows the \vio\ color distribution of
secure+marginal M31 RGB (Class $>0$) and secure+marginal MW dwarf (Class $<0$)
stars.  The color distribution of the M31 RGB stars peaks at \vio\ $\sim 1.5$
with a tail to redder colors.  The MW dwarf stars have a bimodal distribution,
with a peak at \vio\ $\sim 2.3$ and a smaller peak at \vio\ $\sim 1$.  We fit
the color distribution of stars in each of three radial bins as a linear
combination of the RGB and dwarf color distributions: the $R=30$--85~kpc,
$R=85$--125~kpc, and $R=165$~kpc (m11) samples are best fit by 75\%/25\%,
45\%/55\%, and 5\%/95\% RGB/dwarf combinations, respectively.  Stars in our
furthest halo field~m11 closely follow the MW dwarf color distribution, except
for a slight bump at \vio\ $\sim 1.5$.  Such a bump is not present in the
dwarf color distribution, but the M31 RGB color distribution has a peak at
about this color.  This bump in the m11 color distribution is where the 
Class~+3 and all but one of the Class~+1 stars are located.  The colors of
these objects suggest that they are in fact M31 RGB stars.

%%%%%%%%%%%%%%%%%%%%%%%%%%%%%%%%%%%%%%%%%%%%%%%%%%%%%%%%%%%%%%%%%%%%%%%%%%%%%
\subsubsection{Nature of the Bluest Stars (Class~$\pm2$)}\label{sec:bluest}
%%%%%%%%%%%%%%%%%%%%%%%%%%%%%%%%%%%%%%%%%%%%%%%%%%%%%%%%%%%%%%%%%%%%%%%%%%%%%
There are 49~stars in our sample with \xcmd\,$<0$ (i.e.,~bluer than the most
metal-poor isochrone, \vio\,$\lesssim1$).  The \nai, \ki, and TiO diagnostics
have no discriminating power for such blue stars.  Also, these stars lie in
the remote outskirts of both RGB and dwarf PDFs in the CMD and [Fe/H]
diagnostics; the downweighting of outliers in the overall likelihood
calculation ensures that these two diagnostics carry little weight.  This
leaves only two out of the five diagnostics, $L_v$ and $L_{\rm DDO51}$.  As a
further complication, some of the bluest MW dwarfs in our sample also have
very negative velocities that overlap the distribution of M31 RGB stars (see
\S\,\ref{sec:corr_dwf} and Fig.~\ref{fig:clr_vel_gprob}).  For these reasons,
we have taken a close look at these blue stars.

Metal-poor isochrones crowd together in the CMD---e.g.,~an isochrone with
$\rm[Fe/H]=-3$ is only slightly bluer than the bluest ($\rm[Fe/H]=-2.3$)
isochrone in Figure~\ref{fig:cmd} \citep{vdb06}.  Thus any star that lies well
to the blue of our most metal-poor isochrone is almost certainly a MW dwarf
star.  A comparison of the \vio\ color distributions of RGB and dwarf stars
(thin histograms in middle panel of Fig.~\ref{fig:clr_dist}) confirms that
stars with \vio\,$\lesssim1$ are $\approx10\times$ more likely to be MW dwarfs
than M31 RGB stars.  This is reinforced in Figures~\ref{fig:k_diag} and
\ref{fig:tio_diag}: the dwarf PDF contours based on the final training set
extend to \vio\,$\lesssim1$ whereas the RGB contours cut off sharply at
\vio\,$\sim1$.

Of the 49~blue stars, 26 have overall likelihood $\langle{L_i}\rangle<-0.5$
and we feel confident in designating them Class~$-3$ (very secure MW dwarf
stars).  Eleven of the 49 have overall likelihood $\langle{L_i}\rangle>-0.5$
but are significantly bluer than the most metal-poor isochrone
(\xcmd~$<-0.05$); we designate them Class~$-2$ (secure MW dwarf stars).  There
are another 11 stars with overall likelihood $\langle{L_i}\rangle>+0.5$ that
lie slightly blueward of the most metal-poor isochrone ($0.05\le$~\xcmd~$<0$);
they are designated Class~$+2$ (secure M31 RGB stars).  The full set of
classification criteria are given in \S\,\ref{sec:subclass}.

Figure~\ref{fig:class_pm2} shows the distribution of Class~$+2$ and $-2$ stars
in the velocity and \ddo\ diagnostic plots.  The distribution of Class~$+2$
stars in both diagnostics indicates that they are far more likely to be drawn
from the RGB PDF than the dwarf PDF.  The case for the Class~$-2$ stars is
more complicated.  Their \ddo\ parameters span a broad range
[Fig.~\ref{fig:class_pm2}({\bf d})], but this is to be expected: the locus of
blue (\vio\,$\lesssim1$) dwarf stars lies close to the RGB selection box in
the ($M-DDO51$) vs.\ ($M-T_2$) color-color plot \citep[see Fig.~5 of][]{pal03}
so they are sometimes assigned a large \ddo\ parameter value.  The velocities
of Class~$-2$ stars are in the overlap region of the RGB/dwarf PDFs
[Fig.~\ref{fig:class_pm2}({\bf c})].  The Galactic star count model
\citep{bah84,rat85} predicts that the radial velocity distribution of blue MW
dwarf stars in our sample should be skewed toward negative values compared to
that of faint red MW dwarfs \citep[see Fig.~8 of][]{rei02}.  This is because
MW dwarfs with \vio\,$\lesssim1$ must be main-sequence turnoff stars in the MW
halo at heliocentric distances of 15--50 kpc (absolute mag: $(M_I)^{\rm
MSTO}\sim4.0$; apparent mag range of our sample: $20\le I_0\le22.5$).  Members
of the hot, non-rotating MW halo are expected to span a broad range of radial
velocities centered on $-175$\kms\ \citep[reflex of solar motion in the
direction of M31---][]{rei02}, consistent with the observed distribution of
Class~$-2$ stars.

Even though Class~$+2$ and $-2$ stars fall far outside the RGB and dwarf PDFs
in the [Fe/H] diagnostic, it is instructive to compare their \fehp\ and \fehs\
values.  Class~$+2$ stars have median \fehp\ and \fehs\ of $-2.5$ and $-1.9$,
respectively, which are in reasonable agreement given the large measurement
errors for metal-poor stars.  By contrast, Class~$-$2 stars have median \fehp\
and \fehs\ of $\lesssim-4$ and $-2.0$, respectively (the \fehp\ estimates are
admittedly based on gross extrapolation).

In conclusion, the velocities, \ddo\ parameters, and (with greater
uncertainty) [Fe/H] estimates of the Class $+2$ and $-2$ stars are consistent
with their designations as secure M31 RGB and MW dwarf stars, respectively. 

%%%%%%%%%%%%%%%%%%%%%%%%%%%%%%%%%%%%%%%%%%%%%%%%%%%%%%%%%%%%%%%%%%%%%%%%%%%%%
\subsubsection{Nature of the Marginal Cases (Class~$\pm1$)}\label{sec:marg}
%%%%%%%%%%%%%%%%%%%%%%%%%%%%%%%%%%%%%%%%%%%%%%%%%%%%%%%%%%%%%%%%%%%%%%%%%%%%%
The $\langle{L_i}\rangle$ distributions of RGB and dwarf stars are broad
(Fig.~\ref{fig:lhist}): it is likely that some legitimate M31 RGB stars 
have scattered to negative $\langle{L_i}\rangle$ values, and some legitimate 
MW dwarf stars have scattered to positive $\langle{L_i}\rangle$ values.  For
this reason stars with $|\langle{L_i}\rangle|<0.5$ are designated Class +1 or
$-1$ (marginal RGB or dwarf stars; \S\,\ref{sec:subclass}).  

If the shapes of the RGB and dwarf $\langle{L_i}\rangle$ distributions are
the same from field to field, the number of marginal RGB stars (Class~+1) in
a given field can be expressed as the sum of a fraction $a$ of the number of
secure RGB stars (Class~+3 and +2) and a fraction $b$ of the number of secure
dwarf stars (Class~$-3$ and $-2$) in that field:
\begin{equation}\label{eqn:marg_ab}
N({\rm Class}\,+1)~=~a\,[N({\rm Class}\,+3)\,+\,N({\rm
Class}\,+2)]~+~b\,[N({\rm Class}\,-3)\,+\,N({\rm Class}\,-2)].
\end{equation}
Similarly the number of marginal dwarf stars (Class~$-1$) can be expressed
as:
\begin{equation}\label{eqn:marg_cd}
N({\rm Class}\,-1)~=~c\,[N({\rm Class}\,+3)\,+\,N({\rm
Class}\,+2)]~+~d\,[N({\rm Class}\,-3)\,+\,N({\rm Class}\,-2)].
\end{equation}

Figure~\ref{fig:marg}~({\it top\/}) shows the ratio of marginal M31 RGB
(Class~+1) stars to secure RGB and dwarf stars (Class~+3, +2, $-3$, and
$-2$) as a function of the ratio of secure RGB to dwarf stars.  Each data
point represents one of our fields and the abcissa values are roughly in
order of decreasing projected radial distance of the field from the center of
M31.  The data are consistent with $a=0.07$ and $b=0.03$ (solid line) but can
also be fit by $a=b=0.05$ (dashed horizontal line).  The marginal dwarf
(Class~$-1$) stars are well fit by $c=0.03$ and $d=0.15$ [solid line in
Fig.~\ref{fig:marg}~({\it bottom\/})].

Based on the first pair of ($a$,\,$b$) values, about 75\% of the stars in our
overall Class~+1 sample are expected to be true M31 RGB stars while the
remaining 25\% are expected to be MW dwarf stars.  The mix of Class~$-1$ stars
is expected to be 80\% MW dwarfs and 20\% M31 RGB stars.  These percentages
apply to our {\it overall\/} samples of marginal stars; the mix varies from
field to field of course.

Figure~\ref{fig:class_p1} shows the distribution of marginal M31 RGB
(Class~+1) stars in four of our five primary diagnostic plots (all except the
CMD diagnostic).  These stars have radial velocities in the region of overlap
between the RGB and dwarf PDFs, a large spread of \ddo\ values, and \vio\
colors mostly in the range 1--2.5 which places them in the region of overlap
of the RGB and dwarf PDFs in the \nai\ diagnostic plot---all consistent with
their marginal overall likelihood value.  The distribution of these stars in
the [Fe/H] diagnostic plot is consistent with most being M31 RGB stars.

Figure~\ref{fig:class_m1} shows the distribution of marginal MW dwarf
(Class~$-1$) stars in the same four diagnostic plots.  Their \vio\ colors are
mostly in the range 1--2, as for the Class~+1 cases.  The \nai\ diagnostic
plot is not conclusive but the radial velocity, \ddo, and [Fe/H] diagnostic
plots favor the MW dwarf designation.

In summary, the Class~$\pm1$ stars are not as blue as the Class~$\pm2$ stars,
but most of the arguments about the performance of our method on blue stars
applies here as well, only to a lesser degree (\S\,\ref{sec:bluest}).  The
statistical analysis presented above [Fig.~\ref{fig:marg};
Eqns.~(\ref{eqn:marg_ab}--\ref{eqn:marg_cd})] indicates that most of the
Class~+1 objects should be M31 RGB stars and most of the Class~$-1$ objects
should be MW dwarf stars.  The distribution in the diagnostic plots support
this hypothesis.

%%%%%%%%%%%%%%%%%%%%%%%%%%%%%%%%%%%%%%%%%%%%%%%%%%%%%%%%%%%%%%%%%%%%%%%%%%%%%
\subsection{Covariance among Diagnostics}\label{sec:covariance}
%%%%%%%%%%%%%%%%%%%%%%%%%%%%%%%%%%%%%%%%%%%%%%%%%%%%%%%%%%%%%%%%%%%%%%%%%%%%%
%%%%%%%%%%%%%%%%%%%%%%%%%%%%%%%%%%%%%%%%%%%%%%%%%%%%%%%%%%%%%%%%%%%%%%%%%%%%%
\subsubsection{Na, K, and TiO}\label{sec:na_k_tio}
%%%%%%%%%%%%%%%%%%%%%%%%%%%%%%%%%%%%%%%%%%%%%%%%%%%%%%%%%%%%%%%%%%%%%%%%%%%%%
The \nai, \ki\ and TiO diagnostics (Figs.~\ref{fig:na_diag},
\ref{fig:k_diag}, and \ref{fig:tio_diag}) are clearly not independent; not
only do they share a common color axis, but a star's \nai, \ki\ and TiO line
strengths all depend on both temperature (color) and surface gravity.  The
degeneracy between the dwarf and RGB PDFs in these diagnostics at bluer \vio\
colors (especially in \ki\ and TiO) results in much of their discriminating
power coming from the color distribution.  Therefore, if most of the stars in
the sample are bluer than \vio\ $\sim 2.5$, the inclusion of all five of
these diagnostics (\nai, two \ki, and three TiO) will overly weight the color
distribution in determining overall likelihood values.  We choose to use only
the \nai\ diagnostic, since it has the most discriminating power of the five.
However, the \ki\ and TiO diagnostics do have some discriminating power
(beyond the color distribution) for redder stars, and could be used in an
appropriate sample.  Finally, although the two \ki\ diagnostics are obviously
not independent of each other (nor are the three TiO diagnostics), combining
the two \ki\ and three TiO diagnostics is beneficial since this will reduce
the scatter caused by individual line strength measurement errors.  

%%%%%%%%%%%%%%%%%%%%%%%%%%%%%%%%%%%%%%%%%%%%%%%%%%%%%%%%%%%%%%%%%%%%%%%%%%%%%
\subsubsection{Na, CMD, and [Fe/H]}
\begin{figure}
\centerline{\epsfxsize=6in \epsfysize=6in
\epsfbox{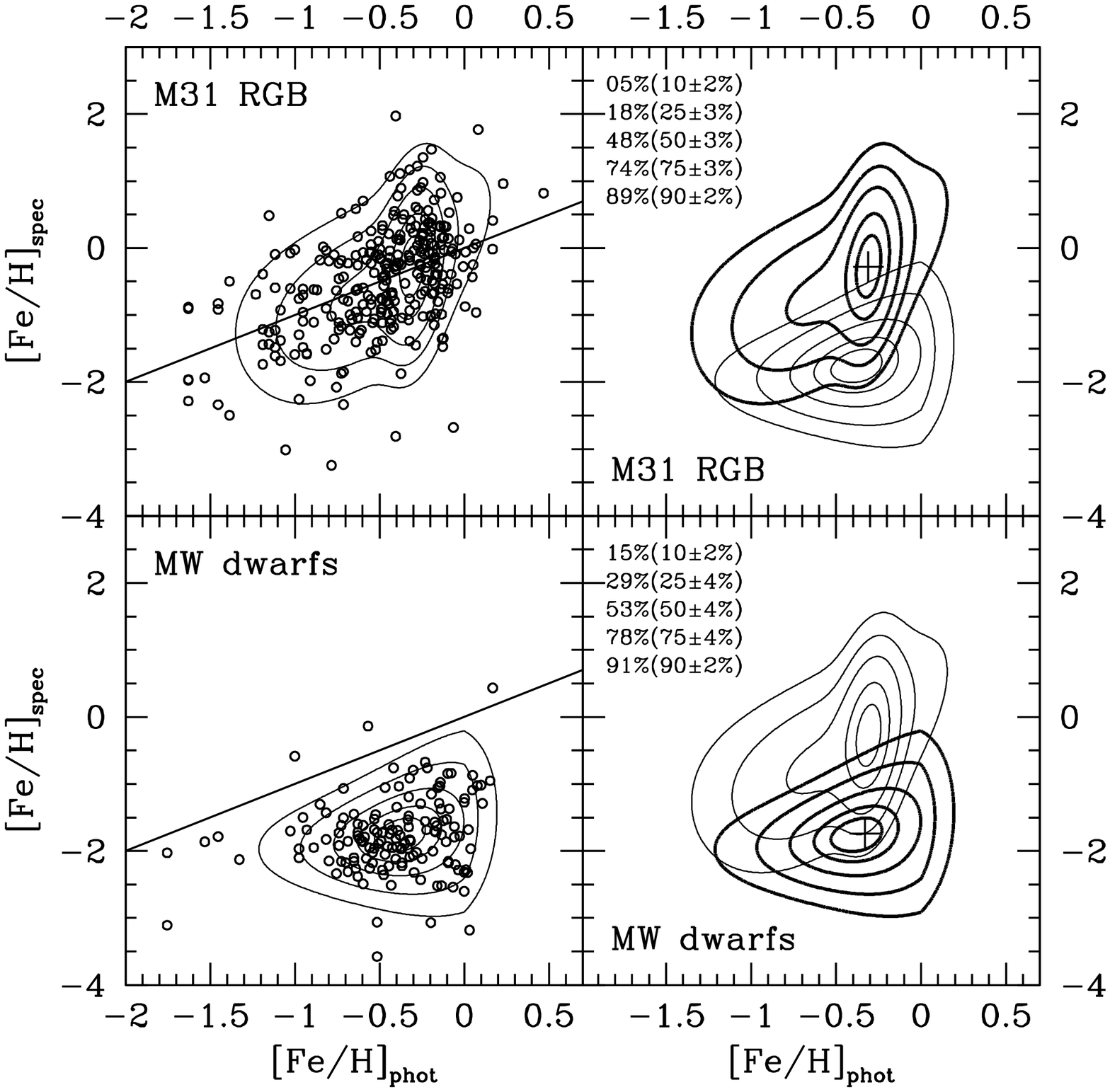}}
\figcaption[fig06.eps]{\label{fig:feh_diag}Same as Figure~\ref{fig:na_diag}
for photometric versus spectroscopic metallicity estimates.  The \fehp\
estimate is derived by comparing the CMD position of each star to 12.6~Gyr
model isochrones with $\rm[\alpha/Fe]=0$ spanning a wide range of
metallicities \citep{vdb06}.  The \fehs\ estimate is derived from the
equivalent width of the \caii\ triplet absorption feature at
$\rm\sim8500~\AA$, using empirical calibration relations based on Galactic
globular cluster red giants \citep{rut97a,rut97b}.  M31 RGB stars show
reasonable agreement between \fehp\ and \fehs, as expected.  The \fehp\
estimate for dwarf stars is obviously incorrect, as it is based on RGB
isochrones at the distance of M31; dwarf stars tend to have relatively weak
\caii\ lines and generally lie well below the \fehp\,=\,\fehs\ relation
(diagonal lines in left panels).
}
\end{figure}

\begin{figure}
\centerline{\epsfxsize=6in \epsfysize=6in
\epsfbox{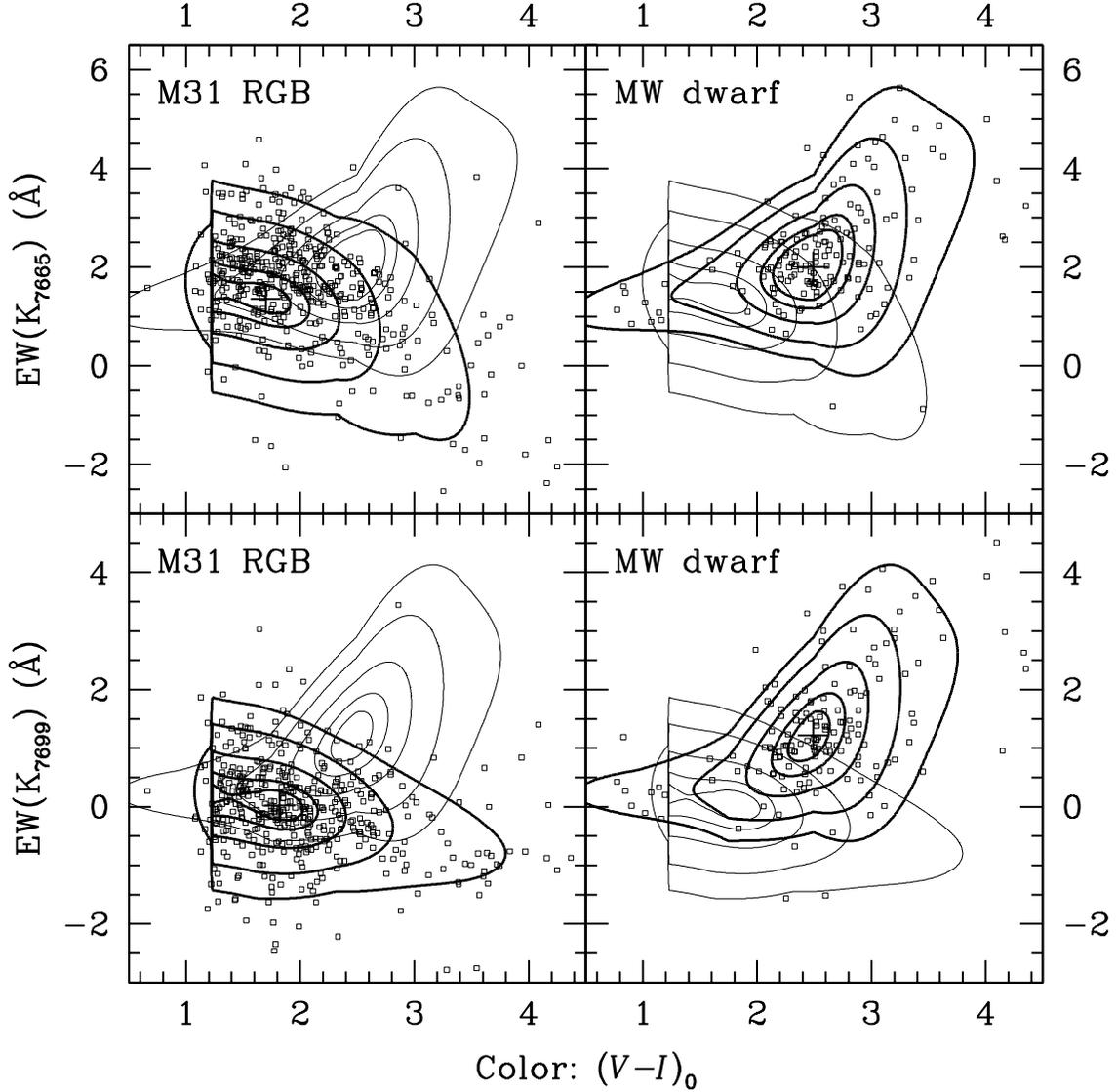}}
\figcaption[fig07.eps]{\label{fig:k_diag}Similar to Figure~\ref{fig:na_diag}
for the \ki\ 7665~\AA\ ({\it top\/}) and 7699~\AA\ ({\it bottom\/}) lines.
Data points show M31 RGB ({\it left\/}) and M31 dwarf ({\it right\/}) stars
from the final training set overlaid on both sets of PDF contours, with the
corresponding set of contours in bold.  There is significant overlap between
the M31 RGB and MW dwarf distributions for bluer colors [\vio\,$>2.5$].
However, the strength of the \ki\ absorption lines increases with increasing
\vio\ for dwarf stars, leading to a clear differentiation between the two
stellar types at redder colors.   
}
\end{figure}

\begin{figure}
\centerline{\epsfxsize=6in \epsfysize=6in
\epsfbox{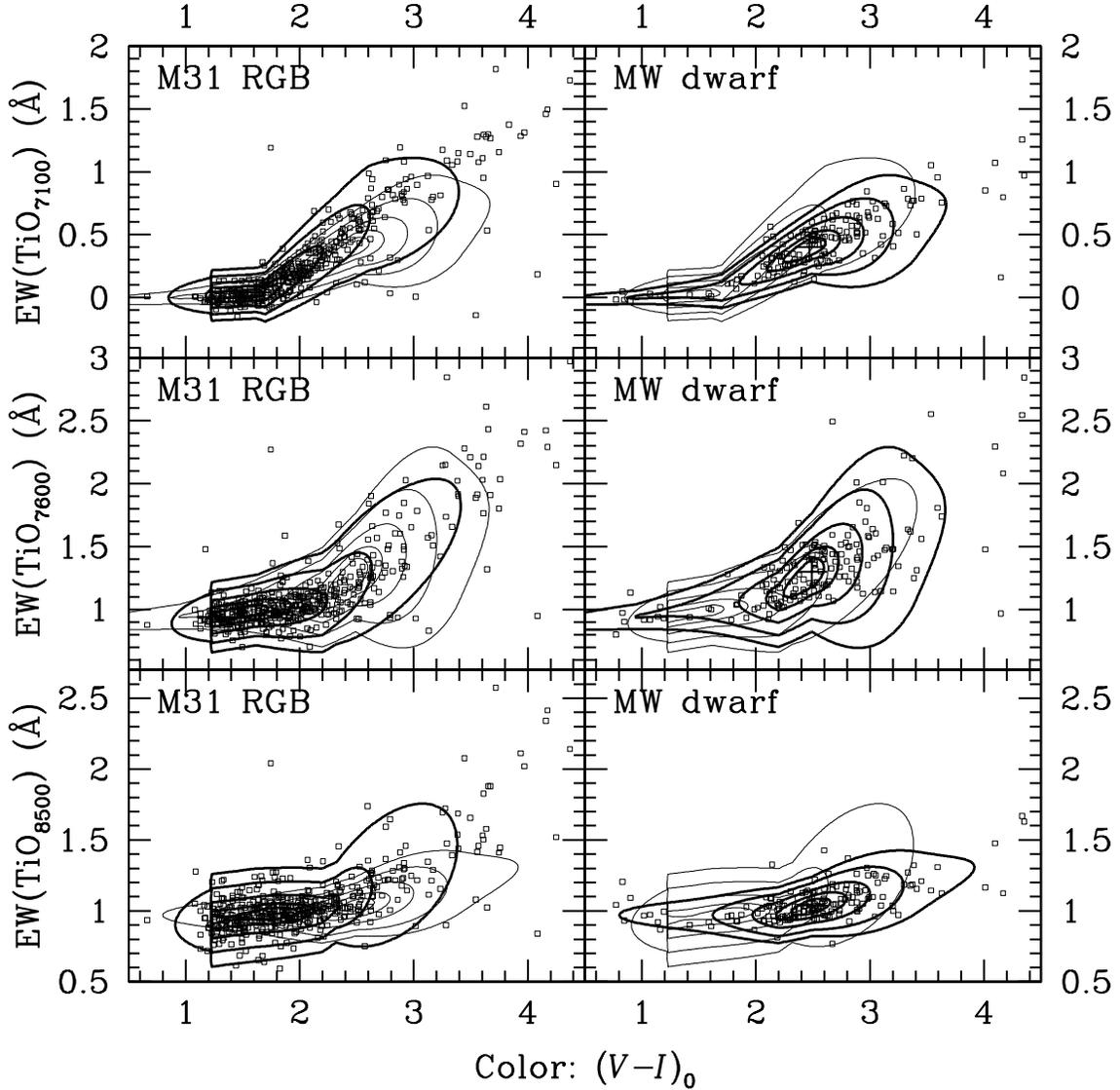}}
\figcaption[fig08.eps]{\label{fig:tio_diag}Same as Figure~\ref{fig:k_diag}
for the TiO 7100, 7600, and 8500~\AA\ bands ({\it top\/}$\rightarrow${\it
bottom\/}), with the RGB and dwarf training set stars and PDFs shown in the
left and right panels, respectively.  There is substantial overlap between
the RGB and dwarf PDFs, especially for the TiO 7600~\AA\ band for which the
measurements are complicated by the presence of the telluric A band.  No TiO
bands are present for relatively hot stars with \vio\ $\la 2.0$, so the
diagnostics provide no RGB/dwarf discrimination in this regime.  However, the
TiO strengths of dwarf and RGB stars follow two slightly different increasing
trends with increasing \vio, so that these diagnostics have some power for
stars redder than \vio\ $\ga 2.0$. 
}
\end{figure}

\begin{figure}
\centerline{\epsfxsize=6in \epsfysize=6in
\epsfbox{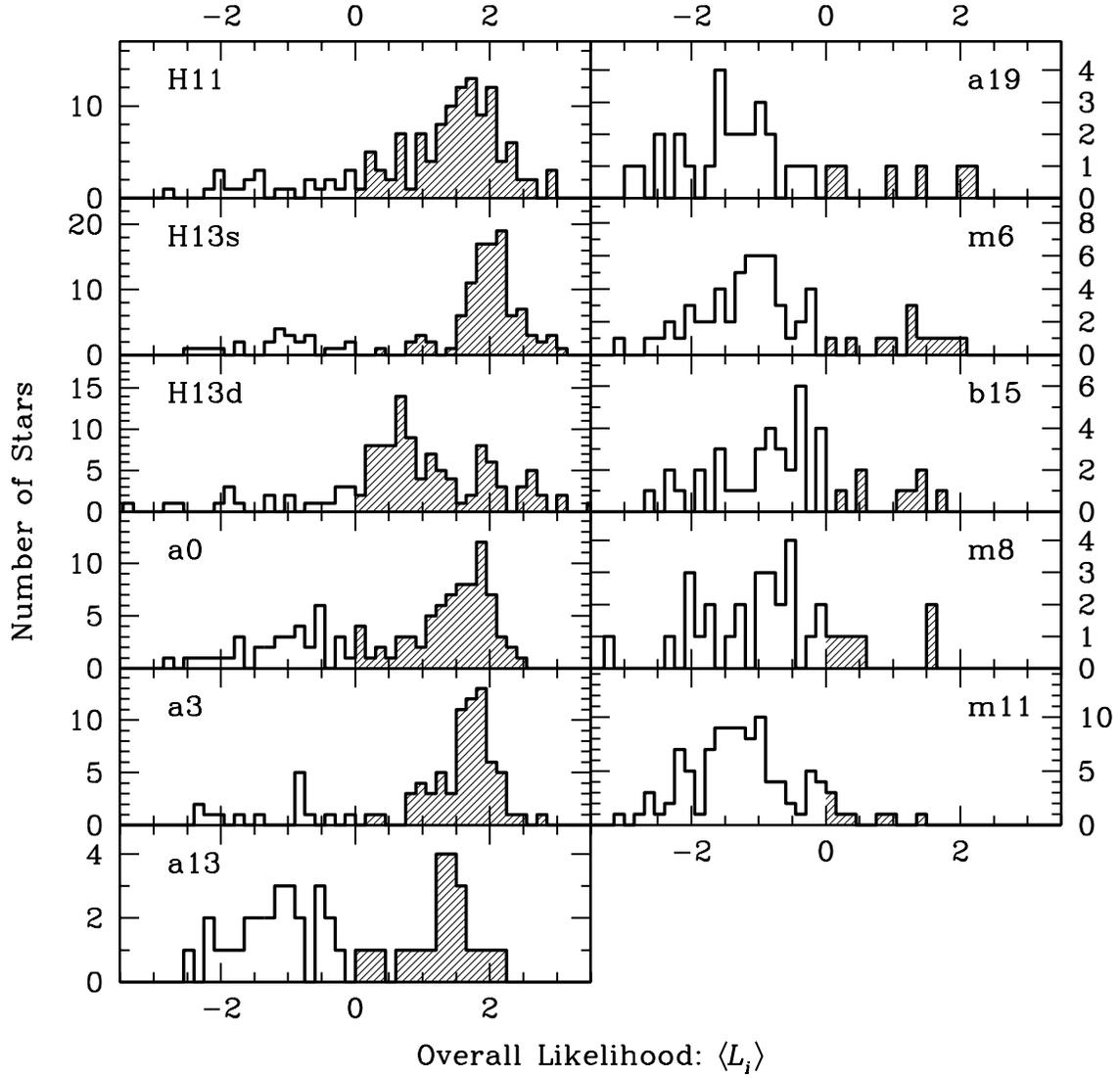}}
\figcaption[fig09.eps]{\label{fig:lhist}Histograms of overall likelihood
($\langle{\it L}_{\it i}\rangle$) values, where $L_i=\log(P_{\rm
giant}/P_{\rm dwarf})$ for each diagnostic, for stars in the M31 outer halo
fields.  The fields are arranged in order of increasing projected distance
from M31 (see Fig.~\ref{fig:map}).  Confirmed RGB stars are those with
$\langle{\it L}_{\it i}\rangle>0$ (shaded histograms) while $\langle{\it
L}_{\it i}\rangle<0$ indicates dwarf stars (open histograms).  There are
clear peaks at $\lesssim-1$ and $\gtrsim+1$, indicating RGB and dwarf star
distributions respectively: in fields with approximately equal numbers of RGB
and dwarf stars, there is a clear bimodality (e.g., fields a0 and a13).  In
inner fields there is a clear RGB peak with a tail of dwarf stars, while in 
outer fields there is a clear dwarf peak with a tail of RGB stars.  As
expected, the red giant fraction decreases with increasing distance from M31.
The exception is field~a3, which due to the increased RGB density in the
southern stream has significantly fewer dwarf stars.  We find secure RGB stars
all the way out to field~m11, which has three distinct outliers (M31 RGB
stars) from the main distribution (Galactic dwarf stars).
}
\end{figure}

\begin{figure}
\centerline{\epsfxsize=5in \epsfysize=6.8in
\epsfbox{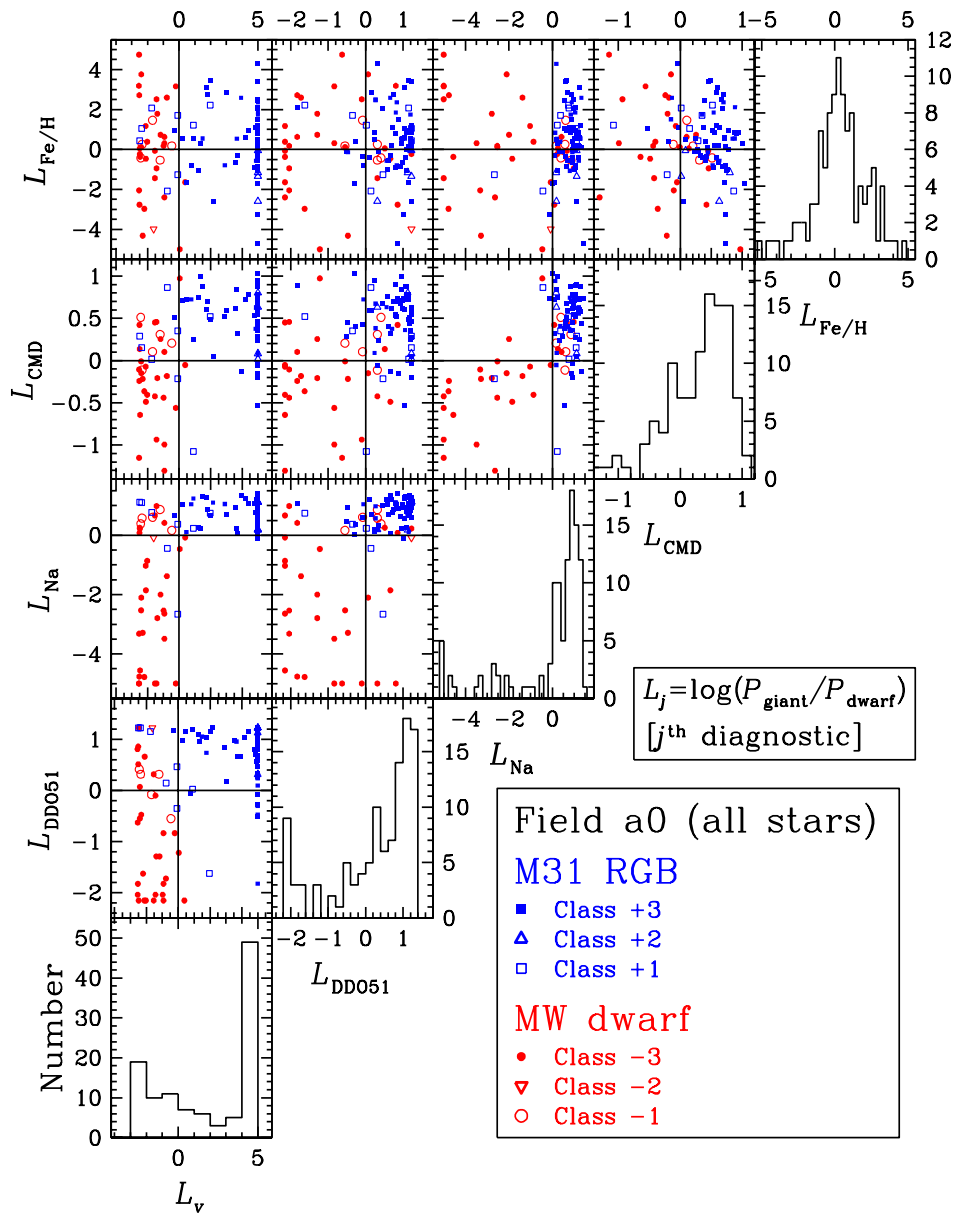}}
\figcaption[fig10.eps]{\label{fig:prob}The position of all stars in field~a0
at $R=30$~kpc in probability ratio space for the five primary diagnostics.
The panels along the right edge diagonal show histograms of $L_j=\log(P_{\rm
giant}/P_{\rm dwarf})$ for each diagnostic $j$; positive values indicate the
star is more likely to be an RGB star than a dwarf (in terms of that
diagnostic) and vice versa for negative values.  The blue and red symbols
indicate M31 RGB stars and MW dwarfs, respectively, as determined by our
likelihood-based method (\S\,\ref{sec:subclass}) and are subdivided into:
Class~$\pm3$ (very secure: filled symbols), Class~$\pm2$ (secure: triangles),
and Class~$\pm1$ (marginal: open squares/circles).  It is reassuring to see
that most RGB stars lie in the top right quadrant of the $L_j$ vs.\ $L_k$
panels, while most dwarf stars lie in the bottom left quadrant.  While no
single diagnostic is a perfect RGB/dwarf discriminant, their combination is
very effective at separating these two populations.
}
\end{figure}

\begin{figure}
\centerline{\epsfxsize=4.5in \epsfysize=6in
\epsfbox{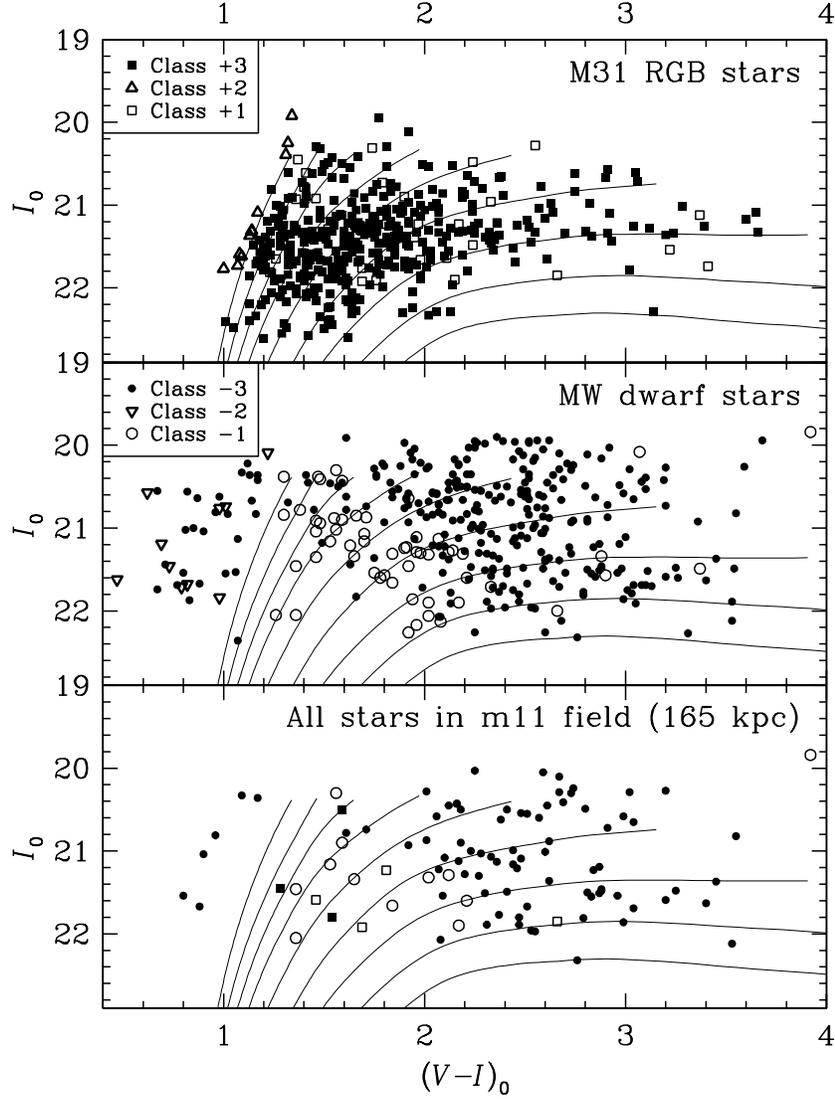}}
\figcaption[fig11.ps]{\label{fig:cmd}Color-magnitude diagram of: ({\it
top\/}) RGB stars from all fields with classes +3, +2, and +1 represented by
filled squares, open triangles, and open squares, respectively (in order of
decreasing certainty of classification; see \S\,\ref{sec:subclass}), ({\it
middle\/}) all dwarf stars with classes $-3$, $-2$, and $-1$ represented by
filled circles, inverted open triangles, and open circles, respectively, and
({\it bottom\/}) all stars in our outermost field~m11 with the same symbols
as above representing the different classes.  The thin curves in each panel
show theoretical RGB tracks from \citet{vdb06} for an age of 12~Gyr,
$\rm[\alpha/Fe]=0$, and metallicities (from left to right) of $\rm[Fe/H]=-2.3$,
$-1.7$, $-1.3$, $-1.0$, $-0.7$, $-0.4$, $-0.1$, $+0.1$, and $+0.4$.
}
\end{figure}

\begin{figure}
\centerline{\epsfxsize=5.5in \epsfysize=5.5in
\epsfbox{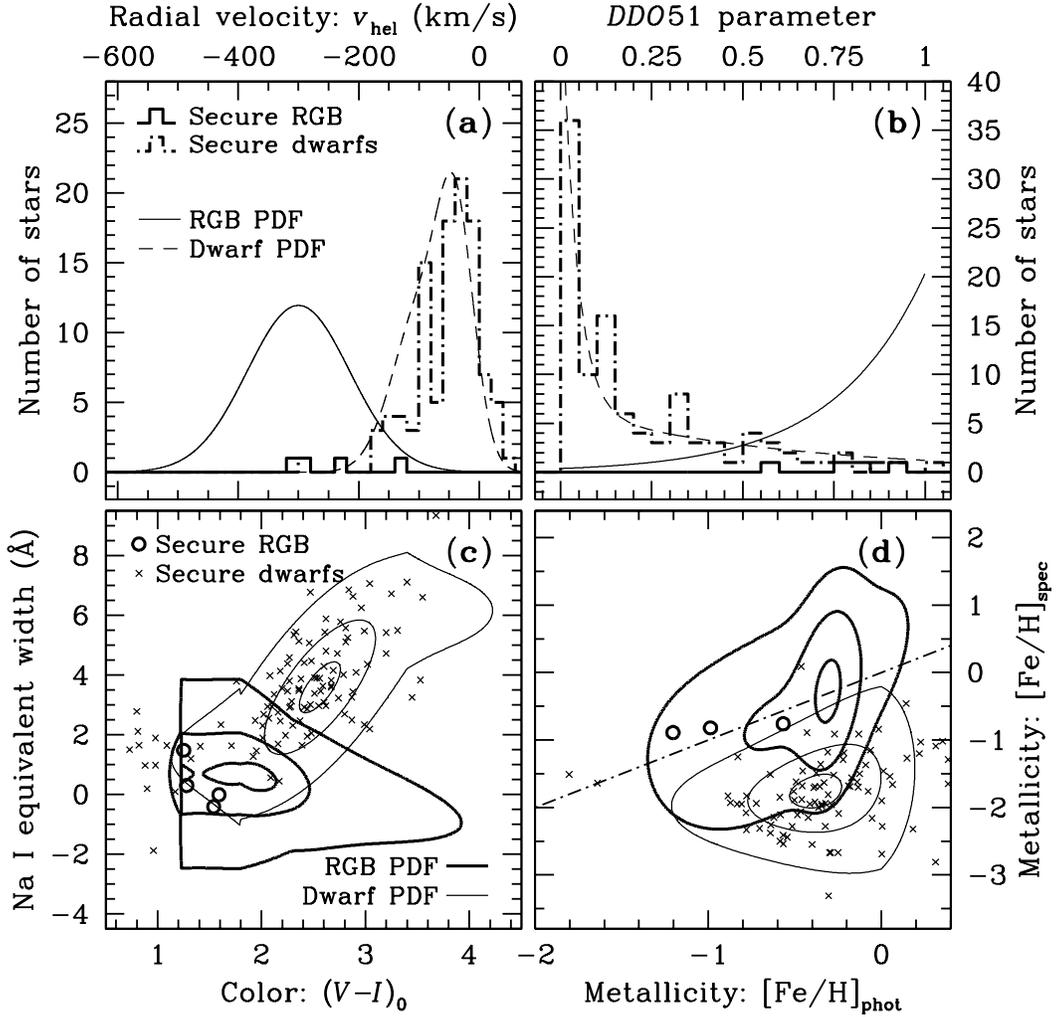}}
\figcaption[fig12.eps]{\label{fig:outrmst_4pnl}({\it a\/})~Radial velocity
histograms for M31 RGB and MW dwarf stars in our two outermost fields~m8
and m11.  Confirmed RGB stars are represented by
the bold solid histogram and secure Galactic dwarf stars by the bold
dot-dashed histogram.  The RGB PDF is shown as a solid curve, and the dwarf
PDF is shown as a dashed curve (Fig.~\ref{fig:vel_diag}).  The stars were
classified as secure RGB or dwarf stars by use of the five-diagnostic
method.  The RGB and dwarf PDFs are normalized to the same area but the
normalization has not been adjusted to match the actual number of stars.~~~
({\it b\/})~Same as ({\it a\/}) for the \ddo\ parameter (see
Fig.~\ref{fig:ddo_diag}).~~~
({\it c\/})~Same as ({\it a\/}) for \vio\ vs.\ \nai\ EW for secure RGB
stars (bold open circles) and dwarfs (crosses) in our two outermost
fields.  The RGB and dwarf two-dimensional PDF contours are shown as bold and
thin lines, respectively (see Fig.~\ref{fig:na_diag}).~~~
({\it d\/})~Same as ({\it c\/}) for \fehp\ vs.\ \fehs\ (see
Fig.~\ref{fig:feh_diag}).  The dot-dashed diagonal line shows the
\fehs\,=\,\fehp\ line.  Only three RGB stars are shown; the fourth is a
significant outlier with respect to both PDFs so this diagnostic recieved
very little weight in the overall likelihood computation for this star
(Eq.~\ref{eqn:weight}).
}
\end{figure}

\begin{figure}
\centerline{\epsfxsize=4.0in \epsfysize=5.5in
\epsfbox{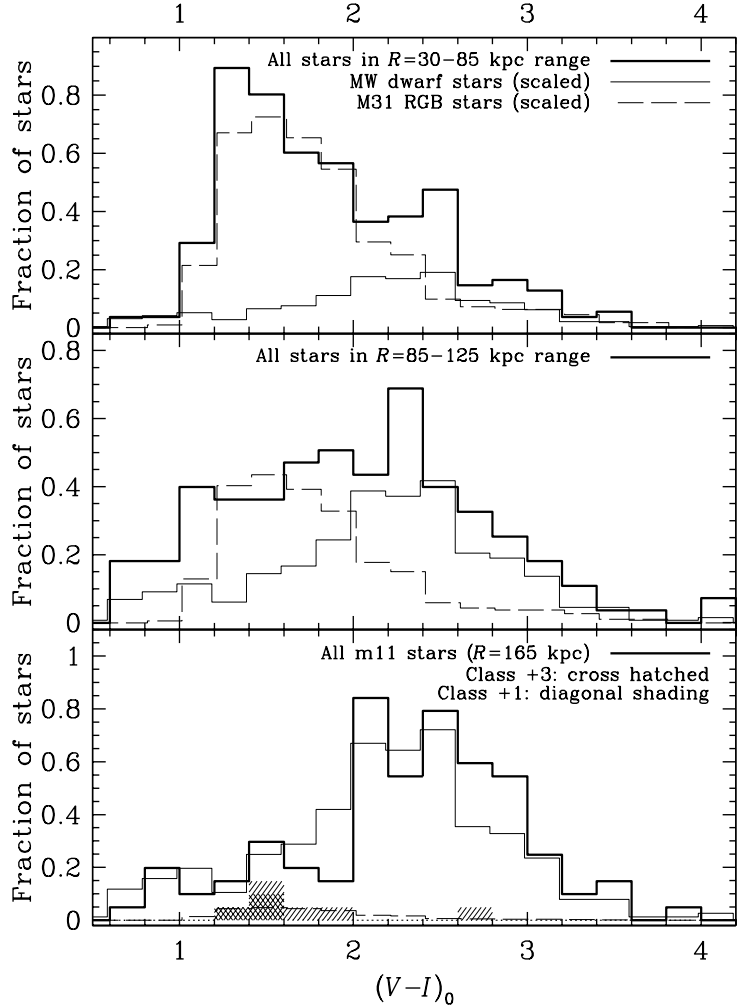}}
\figcaption[fig13.ps]{\label{fig:clr_dist}({\it Top\/})~The \vio\ color
distribution of all stars in fields in the radial range $R=30$--85~kpc (bold
histogram).  This distribution is best fit by a 75\%/25\% combination of the
color distributions of secure+marginal M31 RGB (Class\,$>0$) and MW dwarf
(Class\,$<0$) stars (thin dashed and thin solid histograms, respectively).~~~
({\it Middle\/})~Same as ({\it top\/}) for all stars in fields in the radial
range $R=85$--125~kpc.  The best fit in this case is an approximately equal
mix of M31 RGB and MW dwarf stars, 45\%/55\%.~~~
({\it Bottom\/})~Same as ({\it top\/}) for all stars in our remotest M31 halo
field m11 at $R=163$~kpc.  The m11 stellar \vio\ color distribution generally
follows that of secure+marginal MW dwarf stars.  The \vio\ colors of all three
Class~+3 stars and three of the four Class~+1 stars in m11 (cross-hatched and
diagonally shaded histograms, respectively) coincide with the peak of the
color distribution of secure+marginal M31 RGB stars.  These stars form a slight
apparent peak at \vio\,$\sim1.5$, but the peak is not statistically
significant; in other words, these stars would not have been identified as
M31 RGB stars on the basis of the color distribution alone, but they are
identified by our likelihood method.  A 5\%/95\% combination of M31 RGB and
MW dwarf stars is shown.
}
\end{figure}

\begin{figure}
\centerline{\epsfxsize=6in \epsfysize=6in
\epsfbox{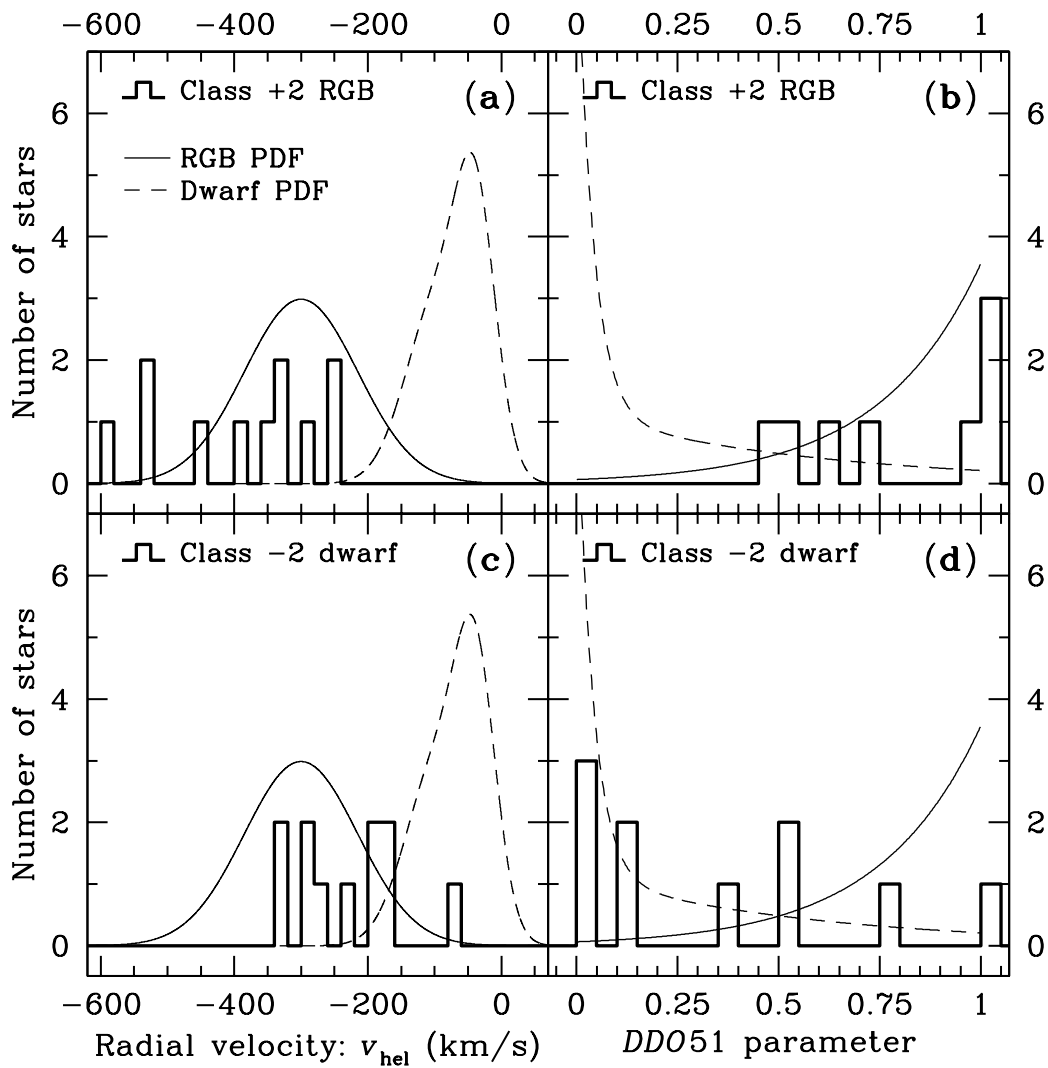}}
\figcaption[fig14.ps]{\label{fig:class_pm2}({\it a\/}--{\it b\/})~Same as
Figure~\ref{fig:outrmst_4pnl}({\it a\/}--{\it b\/}) for Class~+2 M31 RGB
stars as defined in \S\,\ref{sec:subclass} (bold histogram).  The
distribution of radial velocities and \ddo\ parameters for these stars is
a far better match to the RGB PDF (thin solid curve) than to the dwarf PDF
(thin dashed curve).~~~
({\it c\/}--{\it d\/})~Same as ({\it a\/}--{\it b\/}) for Class~$-2$ MW dwarf
stars (bold histogram).
}
\end{figure}

\begin{figure}
\centerline{\epsfxsize=4.5in \epsfysize=6in
\epsfbox{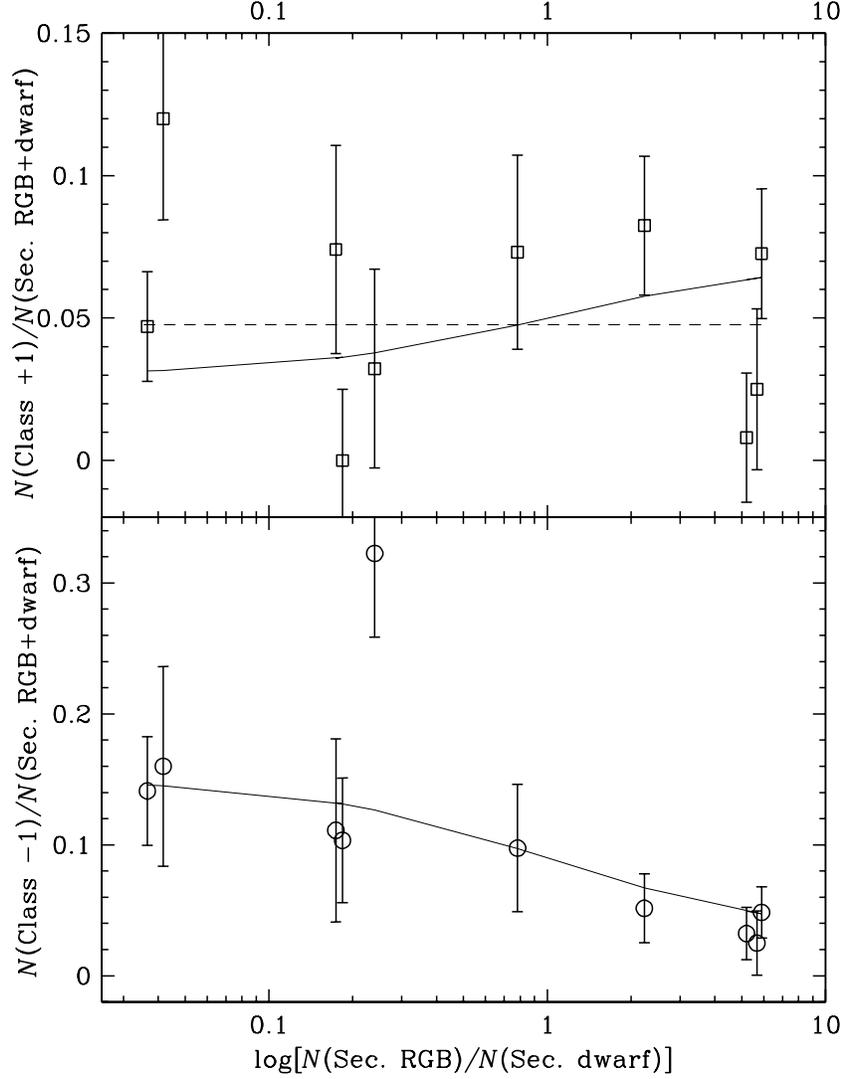}}
\figcaption[fig15.ps]{\label{fig:marg}
({\it Top\/})~Ratio of the number of marginal M31 RGB (Class~+1) stars to the
number of secure RGB+dwarf stars plotted as a function of the ratio of secure
RGB to dwarf stars.  Each data point represents one of our fields (since the
ratio of secure RGB to dwarf stars increases with decreasing radius, the
fields are roughly ordered inversely by radius along the $x$ axis).  The thin
solid curve is a model in which the Class~+1 stars are a specific mix of RGB
and dwarf stars ($a=0.07$ and $b=0.03$; see \S\,\ref{sec:marg}).  The Poisson
error bars shown are based on this model.  An alternate model, $a=b=0.05$, is
shown as a horizontal thin dashed line.~~~
({\it Bottom\/})~Same as ({\it top\/}) for marginal MW dwarf (Class~$-1$)
stars.  The thin solid curve is a model with mix parameters $c=0.03$ and
$d=0.15$.
}
\end{figure}

\begin{figure}
\centerline{\epsfxsize=6in \epsfysize=6in
\epsfbox{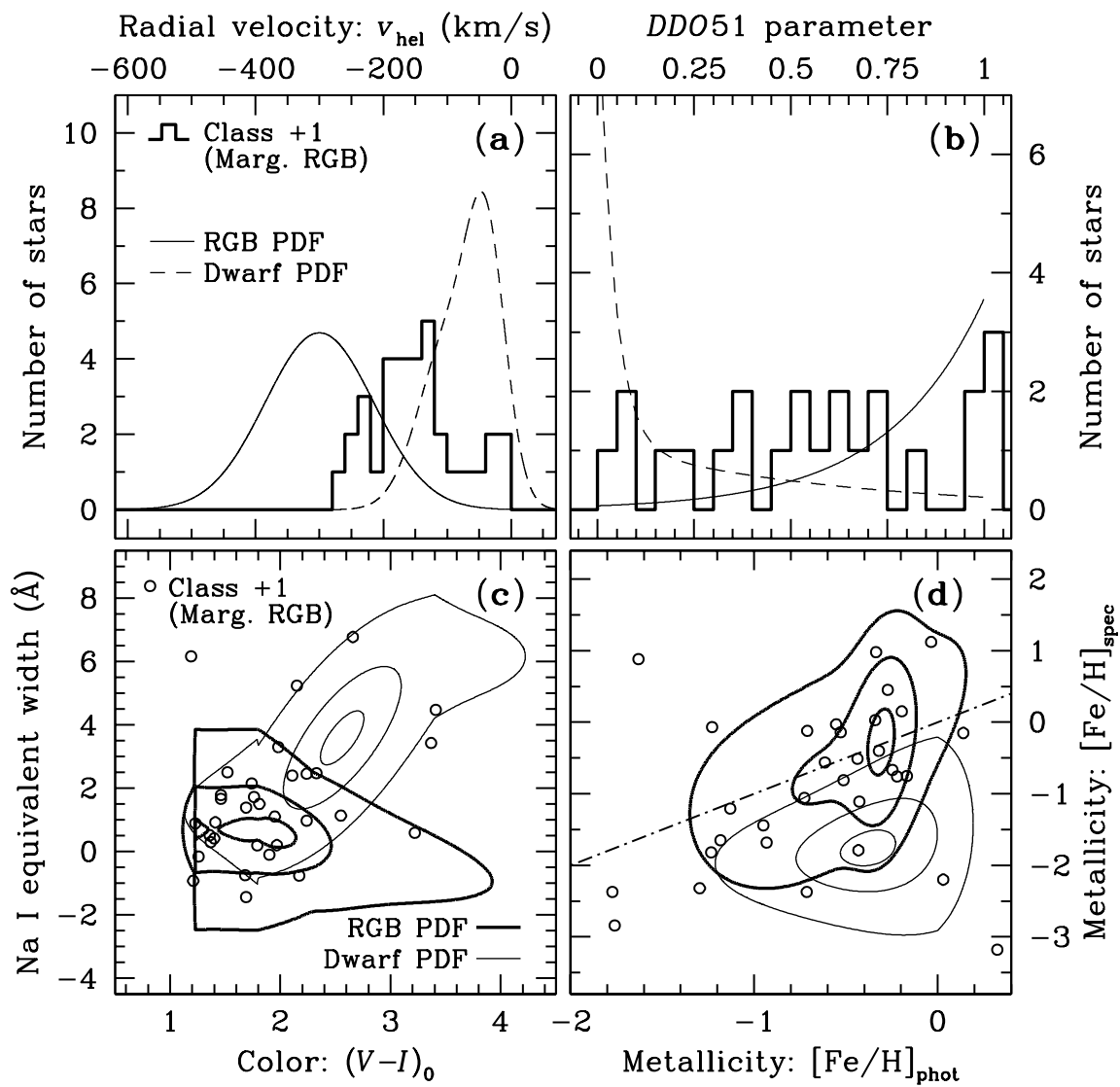}}
\figcaption[fig16.ps]{\label{fig:class_p1}Same as
Figure~\ref{fig:outrmst_4pnl} for marginal M31 RGB (Class~+1) stars
(represented by bold histograms and open circles).  The distribution of this
sample in the four diagnostic plots indicates it consists mostly of M31 RGB
stars with moderately blue colors, \vio~$\sim1$--2.
}
\end{figure}

\begin{figure}
\centerline{\epsfxsize=6in \epsfysize=6in
\epsfbox{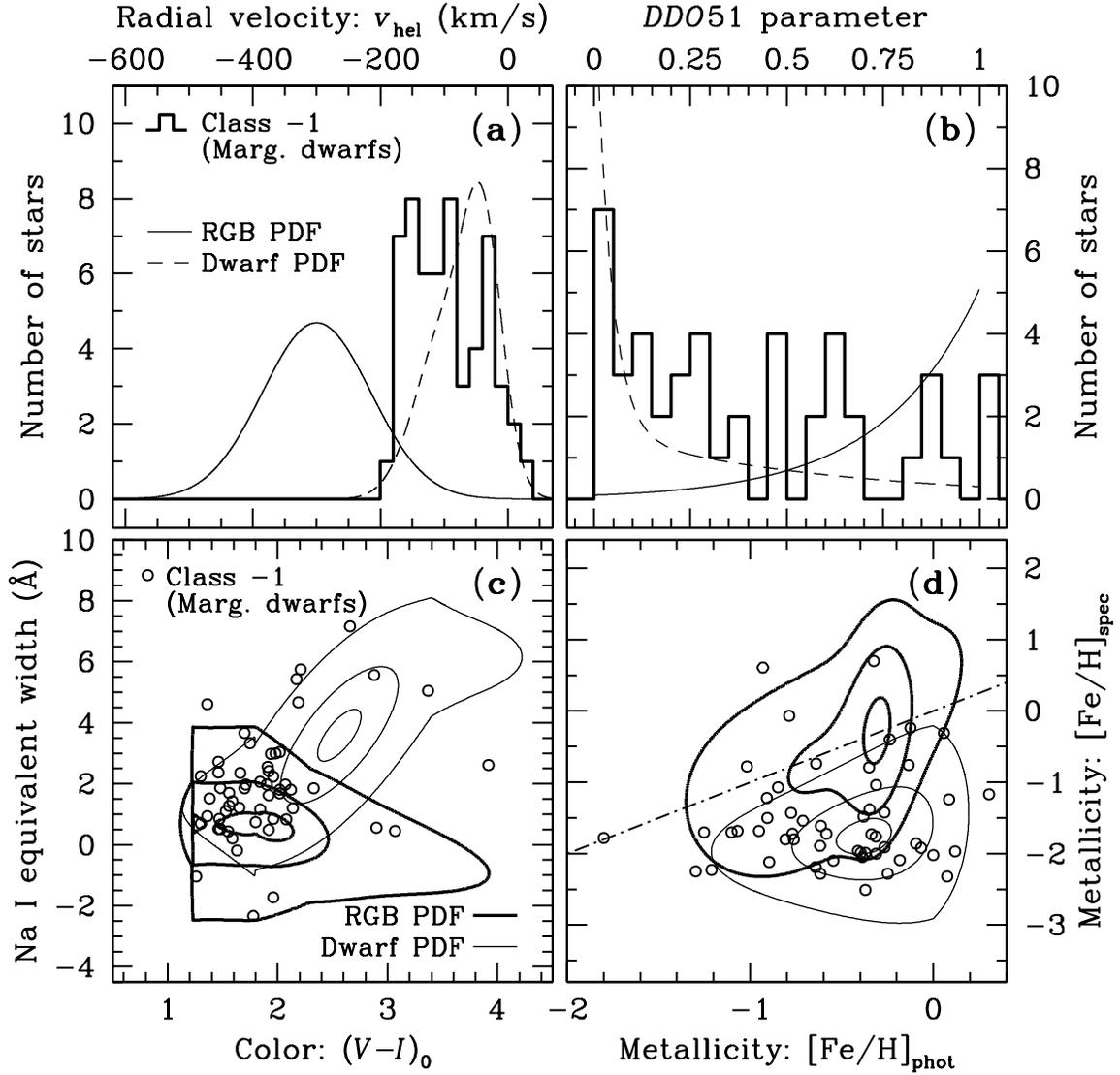}}
\figcaption[fig17.ps]{\label{fig:class_m1}Same as
Figure~\ref{fig:class_p1} for marginal MW dwarf (Class~$-1$) stars
(represented by bold histograms and open circles).  These objects appear to
be mostly MW dwarf stars with moderately blue colors, \vio~$\sim1$--2,
judging from their distribution in the four diagnostic plots.
}
\end{figure}

\begin{figure}
\centerline{\epsfxsize=4.5in \epsfysize=6in
\epsfbox{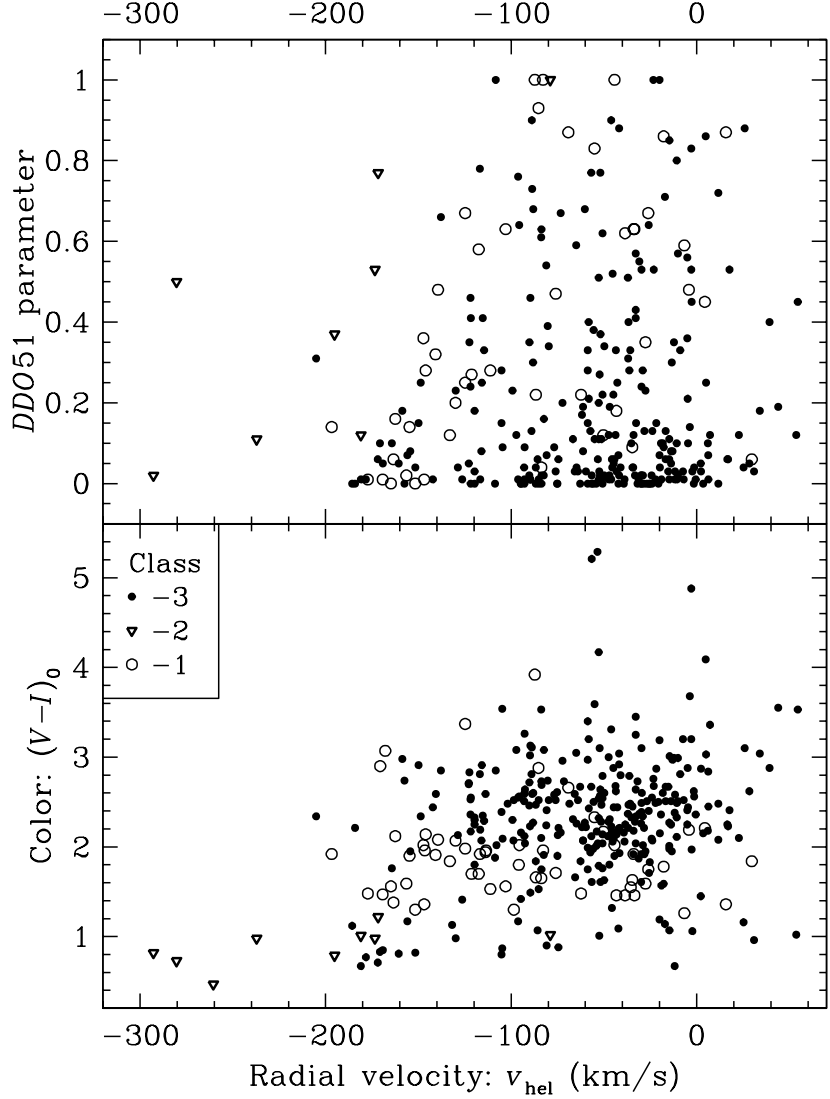}}
\figcaption[fig18.ps]{\label{fig:clr_vel_gprob}The \ddo\ parameter ({\it
top\/}) and \vio\ color ({\it bottom\/}) plotted versus radial velocity for
secure MW dwarf stars.  As expected, dwarf stars tend to have small values
of the \ddo\ parameter but there is a tail to the distribution all the way up
to values near unity.  No significant/strong trend is seen in velocity vs.\
\ddo.  There is a slight trend in the velocity vs.\ \vio\ plot in the sense
that the handful of dwarf stars with the most negative velocities are all
very blue; the rest of the dwarf stars ($v\gtrsim-150$~\kms) span a large
range of \vio\ colors.
}
\end{figure}

%%%%%%%%%%%%%%%%%%%%%%%%%%%%%%%%%%%%%%%%%%%%%%%%%%%%%%%%%%%%%%%%%%%%%%%%%%%%%
Although the vertical axes of the \nai, CMD, and [Fe/H] diagnostics 
(Figs.~\ref{fig:na_diag}, \ref{fig:cmd_diag}, and \ref{fig:feh_diag}) are
independent of each other, there is a complicated dependence between the 
horizontal axes of these three diagnostics.  The horizontal axis of the \nai\ 
diagnostic is \vio\ color.  Both the horizontal axes of the CMD and [Fe/H] 
diagnostics are dependent on color: \xcmd\ depends on position within the CMD
with relation to the theoretical RGB isochrones (\S\,\ref{sec:cmd}), as does
\fehp\ (\S\,\ref{sec:feh}).  However, the dwarf and RGB PDFs in both the CMD
and [Fe/H] diagnostics cover the same horizontal range.  All the
discriminating power in these two diagnostics comes from the vertical axes,
which are independent. 

%%%%%%%%%%%%%%%%%%%%%%%%%%%%%%%%%%%%%%%%%%%%%%%%%%%%%%%%%%%%%%%%%%%%%%%%%%%%%
\subsubsection{Correlations within the Dwarf Star Sample}\label{sec:corr_dwf}
%%%%%%%%%%%%%%%%%%%%%%%%%%%%%%%%%%%%%%%%%%%%%%%%%%%%%%%%%%%%%%%%%%%%%%%%%%%%%
We have explored all secure M31 dwarf stars for possible correlations between
radial velocity and \vio\ color and/or \ddo\ parameter
(Fig.~\ref{fig:clr_vel_gprob}).  Most dwarf stars have small \ddo\ parameters,
as expected.  Moreover, regardless of radial velocity, there is a tail to the
distribution of \ddo\ parameters reaching all the way up to values near unity.
No strong trend is seen between radial velocity and \vio\ color either.  The
few dwarf stars with very negative radial velocities ($v_{\rm
hel}<-225$~\kms) are all very blue.  However, the rest of the dwarf stars
display a wide range of \vio\ colors and show no correlation between radial
velocity and color.

\begin{figure}
\centerline{\epsfxsize=6in \epsfysize=6in
\epsfbox{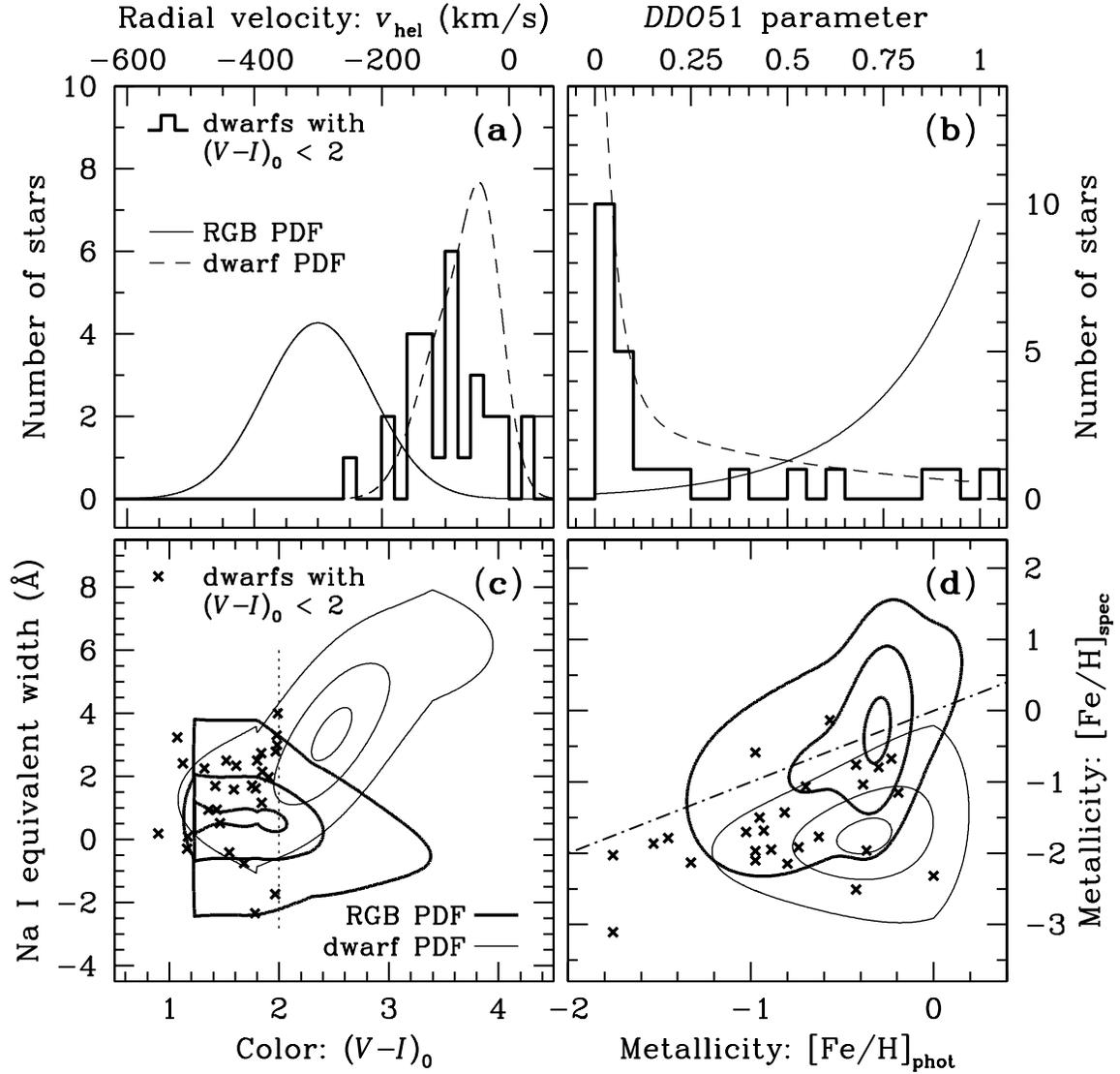}}
\figcaption[fig19.ps]{\label{fig:ind_test}Same as Figure~\ref{fig:outrmst_4pnl}
for stars in the foreground Galactic dwarf training set with $(V-I)_0<2$
(represented by the bold histograms and crosses).  This color cut, shown as a
vertical dotted line in panel~({\it c\/}), is designed to isolate the tail of
the training set dwarf distribution in the portion of the \vio\ vs.\ \nai\ EW
plane where the RGB and dwarf PDFs overlap.  Even though the stars shown in
this figure are drawn from the tail of the \vio\ distribution, they do not
show a strong bias relative to the overall dwarf distribution in the radial
velocity and \ddo\ diagnostic plots.  There is obviously a correlation
between \vio\ and \fehp\ (latter being based on CMD position); as a result
the dwarfs in panel~({\it d\/}) display a bias relative to the dwarf
PDF in the \fehp\ vs.\ \fehs\ plane.
}
\end{figure}

To ensure that the stars we have identified as secure M31 RGB stars are not
simply the tail of the dwarf distribution in the outermost, sparsest M31 halo
fields, we explored the dwarf training set for correlations between the
positions of stars in the diagnostics.  If a star in the tail of the dwarf
distribution in one diagnostic is also in the tail of the dwarf distribution
in the other diagnostics, it will receive a low probability of being a dwarf
star in all of them, adding up to an spuriously high probability of being an
RGB star.  For four of the most powerful diagnostics (radial velocity, \nai\
EW vs.\ \vio\ color, \ddo\ parameter, and \fehp\ vs.\ \fehs), we chose dwarf
training set stars in the tail of the dwarf distribution and investigated
where they fell in the other three diagnostics.  Figure~\ref{fig:ind_test}
shows where stars selected to be in the tail of the dwarf distribution in the
\nai\ diagnostic (\vio\ $< 2$) fall in the other diagnostics.  It is clear
that they do not fall in the tail of the dwarf distribution in radial
velocity, \ddo\ parameter, or in \fehs\ vs.\ \fehp.  The only correlation is a
slight shift to lower \fehp\ values.  This is expected, since \fehp\ is
related to \vio\ color.  The tails of the dwarf distribution in the other
diagnostics show the same behavior as in this example: no correlation was
found between diagnostics.

%%%%%%%%%%%%%%%%%%%%%%%%%%%%%%%%%%%%%%%%%%%%%%%%%%%%%%%%%%%%%%%%%%%%%%%%%%%%%
\subsection{Shortcomings of the Present Method and Future
Improvements}\label{sec:future}
%%%%%%%%%%%%%%%%%%%%%%%%%%%%%%%%%%%%%%%%%%%%%%%%%%%%%%%%%%%%%%%%%%%%%%%%%%%%%
While our likelihood-based RGB/dwarf separation method is vastly superior to
the radial velocity cut used in other spectroscopic surveys of M31 stars
(\S\,\ref{sec:rvcut}), it is far from optimal and will no doubt undergo
improvements in the not-too-distant future.  These improvements are beyond
the scope of the present paper and will be tackled in a future paper; we list
them here for the sake of completeness.

%%%%%%%%%%%%%%%%%%%%%%%%%%%%%%%%%%%%%%%%%%%%%%%%%%%%%%%%%%%%%%%%%%%%%%%%%%%%%
\subsubsection{Color Dependence of the Diagnostics}\label{sec:ind_diag}
%%%%%%%%%%%%%%%%%%%%%%%%%%%%%%%%%%%%%%%%%%%%%%%%%%%%%%%%%%%%%%%%%%%%%%%%%%%%%
The covariance of the diagnostics has been discussed in
\S\,\ref{sec:covariance}.  The degeneracy between the horizontal axes
(color-dependence) of the two-dimensional diagnostics can be lifted by
converting them to one-dimensional diagnostics.  
One method which holds promise for the future is to fit a spline to the 
distribution of training set stars in each diagnostic.  The distribution of 
vertical distances of the training set stars
from the fit will yield a one-dimensional PDF.  Once a two-dimensional 
diagnostic is collapsed into a one-dimensional diagnostic in this way, the
latter will have no sensitivity to differences in the \vio\ color
distribution between M31 RGB and MW dwarf star populations.  This will enable
us to incorporate the \ki\ and TiO diagnostics into the overall likelihood.

%%%%%%%%%%%%%%%%%%%%%%%%%%%%%%%%%%%%%%%%%%%%%%%%%%%%%%%%%%%%%%%%%%%%%%%%%%%%%
\subsubsection{M31's Radial Metallicity Gradient}\label{sec:rad_trnd}
%%%%%%%%%%%%%%%%%%%%%%%%%%%%%%%%%%%%%%%%%%%%%%%%%%%%%%%%%%%%%%%%%%%%%%%%%%%%%
Another future improvement to the method will be to account for radial trends
in the M31 RGB star properties.  We are currently using one M31 RGB training 
set for all of our fields.  However, we have discovered that M31 RGB stars at
large radial distances are systematically more metal-poor than those at
smaller radial distances \citep{kal06a}.  Our present method is sub-optimal
in that our RGB PDFs do not track the change in the \vio\ color distribution
of M31 stars with radial distance and are based on a training set that is
drawn mostly from M31's inner spheroid.  In order to track this radial trend,
we will need a significantly larger sample of RGB stars in M31's outer halo
to use as a training set.  We expect that a much larger outer halo RGB sample
will be available in the next few years.

%%%%%%%%%%%%%%%%%%%%%%%%%%%%%%%%%%%%%%%%%%%%%%%%%%%%%%%%%%%%%%%%%%%%%%%%%%%%%
\subsubsection{Accounting for Measurement Error and Eliminating Analytic
PDFs}\label{sec:gauss_err}
%%%%%%%%%%%%%%%%%%%%%%%%%%%%%%%%%%%%%%%%%%%%%%%%%%%%%%%%%%%%%%%%%%%%%%%%%%%%%
The likelihood-based method developed in this paper does not explicitly take
advantage of our (admittedly rough) knowledge of the measurement errors in
the quantities used in our various diagnostic plots (\S\,\ref{sec:error}).
For idealized error distributions, each training set star can be treated as a
Gaussian whose width is the quadrature sum of its own measurement error and
that of the star for which the likelihood estimate is being made.  Naturally,
these would be one-dimensional Gaussians for the radial velocity and \ddo\
diagnostics and two-dimensional Gaussians for the \nai, CMD, and [Fe/H]
diagnostics.  The integral under the Gaussian would be the same for each
training set star.  The M31 RGB and MW dwarf PDFs would then be the sum of
all the Gaussians corresponding to the training set stars.  The integral
under each PDF would be set to unity.  We hope to implement such a procedure
in the future; with enough stars in the training set the sum of all the
Gaussians should be smooth enough that it will no longer be necessary to fit
analytic functions in order to derive PDFs.

%%%%%%%%%%%%%%%%%%%%%%%%%%%%%%%%%%%%%%%%%%%%%%%%%%%%%%%%%%%%%%%%%%%%%%%%%%%%%
\subsubsection{Estimating the Number of M31 RGB Stars}\label{sec:num_rgb}
%%%%%%%%%%%%%%%%%%%%%%%%%%%%%%%%%%%%%%%%%%%%%%%%%%%%%%%%%%%%%%%%%%%%%%%%%%%%%
Through most of this paper we have focused on the question: Which stars are
M31 RGB stars and which ones are MW dwarf stars?  Alternatively, one might
ask: What is our best statistical estimate of the number of M31 RGB stars and
MW dwarf stars in the sample?  The first question is clearly important for
addressing a wide variety of scientific topics (such as the metallicity or
dynamics of a clean sample of M31 RGB stars), but the second question is more
relevant when it comes to studies of the spatial density distribution of RGB
stars and surface brightness profile and M31's halo.

The distinction between the above two questions can be illustrated by the
following simple scenario.  Let us assume there is a measurable $x$ and that
M31 RGB stars and MW dwarf stars are characterized by different, but
overlapping $x$ distributions.  Armed with only this information, it is
impossible to assign a clear designation to any star whose $x$ value lies in
the overlap region.  However, if the {\it shape\/} of underlying parent $x$
distribution is known for M31 RGB and MW dwarf stars, one can fit a weighted
sum of the RGB and dwarf star parent distributions to the observed $x$
distribution and thereby determine the fraction of each stellar type in the
sample.

In the context of our method, the RGB and dwarf PDFs for each diagnostic
(modulo the other planned improvements of course) can be treated as the
underlying parent distributions.  One could then fit, simultaneously for all
diagnostics, a weighted sum of the PDFs to the observed distribution of
points.  The outcome of the fit would be our best estimate for the M31 RGB
fraction in each field.  We plan to explore such a method in a future paper.

%%%%%%%%%%%%%%%%%%%%%%%%%%%%%%%%%%%%%%%%%%%%%%%%%%%%%%%%%%%%%%%%%%%%%%%%%%%%%
\subsection{The Diagnostic Method Versus a Radial Velocity
Cut}\label{sec:rvcut}
%%%%%%%%%%%%%%%%%%%%%%%%%%%%%%%%%%%%%%%%%%%%%%%%%%%%%%%%%%%%%%%%%%%%%%%%%%%%%
We assess the performance of the likelihood-based method presented in
this paper by comparing it to the use of a simple radial velocity cut to
select M31 RGB stars.  Most previous spectroscopic studies of M31 RGB stars
have resorted to using radial velocity cuts: for example, \citet{rei02} used
$v<-220$~\kms\ while \citet{iba05} used $v<-100$~\kms.

To investigate the improvement in M31 RGB sample size gained through using
our diagnostic method, we applied a velocity cut of $v<-220$~\kms\ to our
smooth spheroid sample.  (The disk-dominated field H13d and the stream stars in
fields H13s and a3 are excluded from this count.)  This resulted in an RGB
sample of 208 stars, compared to 207 stars identified by the likelhood method
as secure (Class~+3 and +2) RGB stars.  However, only 192 stars are in
common between the two samples.  This velocity cut recovers 93\% 
of the M31 RGB stars, and 8\% of the velocity-selected RGB stars are actually
MW dwarf star contaminants.  If we restrict our analysis to
the outer halo fields (fields a13 through m11), this velocity cut yields 42
stars while our likelihood method yields 40 secure RGB stars, with 36
stars in common.  Thus, a velocity cut of $v<-220$~\kms\ in the outer halo 
recovers 90\% of the M31 RGB stars in these fields, but suffers from a MW 
dwarf star contamination rate of 19\%.  As an extreme case, an $L_v>+0.5$
cut in the radial velocity diagnostic (which corresponds $v<-190$~\kms) yields
four stars in our outermost field~m11 ($R=165$~kpc), but only two of these are
secure M31 RGB stars, while one each is a secure MW dwarf and a marginal MW
dwarf---i.e.,~a 50\% contamination rate.  Moreover, one of the secure M31 RGB
stars in this field fails this velocity cut.

If a less conservative velocity cut is made, all the RGB stars can be
recovered, but at the price of a larger dwarf contamination rate.  When a
velocity cut of $v<-100$ \kms\ is made in H11, our innermost smooth spheroid
field, it yields 120 stars, while our likelihood method finds 102 secure 
RGB stars.  The sample selected solely on the basis of 
this velocity cut would have a dwarf contamination rate of 15\%.  If this
velocity cut is made in field a0 as well as H11, it yields 203 stars, versus
167 secure RGB stars, for a dwarf contamination rate of 17.7\%.  Finally, if 
a somewhat more conservative velocity cut of $v<-160$ \kms\ \citep[as
in][]{cha06} is made in the smooth halo fields beyond $R=30$~kpc from the
nucleus (a0, a13 through m11) the total number of stars selected is 140, 
compared to 105 secure RGB stars, for a dwarf contamination rate of 25\%.

In conclusion, we find that the radial velocity cuts that have been used
to isolate M31 RGB stars in other spectroscopic surveys
\citep[e.g.,][]{rei02,iba05,cha06} yield samples that contain a significant
fraction of foreground MW dwarf star contaminants and/or are incomplete.
While this may be acceptable for statistical studies of the inner regions of
M31 where the surface density of RGB stars greatly exceeds that of foreground
MW dwarf stars, the dwarf contamination presents a serious obstacle for
studies of M31's sparse outer halo.  As discussed above, using a conservative
radial velocity cut (say $-200$\,\kms)
reduces the dwarf contamination rate but causes other complications.
For example, any dynamical analysis of a radial velocity-selected sample of
RGB stars would have to take the cut into account.  Given the large spread
of M31 RGB radial velocities around $v_{\rm sys}^{\rm M31}=-300$\,\kms\ (in
its dynamically hot bulge and halo and fast rotating disk), a radial velocity
cut of $-200$\,\kms\ would lead to substantial incompleteness/bias.  By
contrast, our likelihood-based method produces a clean unbiased sample of M31
RGB stars that can directly be used for dynamical (and other) studies.

%%%%%%%%%%%%%%%%%%%%%%%%%%%%%%%%%%%%%%%%%%%%%%%%%%%%%%%%%%%%%%%%%%%%%%%%%%%%%
\subsection{Comparison to Photometric Studies}\label{sec:comp_phot} 
%%%%%%%%%%%%%%%%%%%%%%%%%%%%%%%%%%%%%%%%%%%%%%%%%%%%%%%%%%%%%%%%%%%%%%%%%%%%%
In this section we switch our attention from comparison with other
spectroscopic studies to comparisons with photometric studies of the surface
brightness profile of M31.  It has long been recognized that star count
studies are far superior to integrated light measurements
\citep{pri87,pri88,pri94}.  Star count studies typically select M31 RGB
candidates based on location within the CMD and use control fields for
statistical subtraction of the ``background'' \citep[i.e., foreground MW
dwarf star and background field galaxy contaminants---][]{pri94,dur04,irw05}.
However, systematic errors in background subtraction (e.g., due to
photometric and/or seeing variations) limit these techniques to regions of
M31 where the surface density of RGB stars is comparable to or greater than
the background level ($\mu_V\approx30$~mag~arcsec$^{-2}$).

A variety of photometric filtering techniques have been used to increase the
contrast of M31 RGB stars with respect to contaminants.  An example of this
is the use of $UBRI$ photometry to reject background galaxies \citep*{rei98}.
As discussed in \S\,\ref{sec:data}, \citet{ost02} used \ddo\ photometry to
select M31 RGB candidates.  This technique reduces the surface brightness
level of the background to $\mu_V\gtrsim31$~mag~arcsec$^{-2}$ \citep{guh05}.

In contrast to the above methods, using a combination of photometric and
spectroscopic diagnostics allows us to reliably pick out very sparse groups
of M31 RGB stars by rejecting foreground MW dwarf star and background field
galaxy contaminants.  The three secure M31 RGB stars identified in our
outermost field m11 constitute only a few percent of the overall sample of
\ddo-selected RGB candidates in that field; the rest are MW dwarf stars or
galaxies.  We use the ratio of secure M31 RGB stars to secure MW dwarf
stars, in conjunction with a Galactic model \citep{bah84,rat85}, to estimate
M31's surface brightness profile \citep{guh05}.  This allows us to map the
outer halo of M31 down to an unprecedented surface brightness level of
$\mu_V\sim35$~mag~arcsec$^{-2}$.  \citet{kal06a} study the metallicity
distribution of M31 RGB stars isolated by our diagnostic method and find that
the stellar halo is significantly more metal poor than the bulge; this is in
contrast to findings from previous photometric studies of a metal-rich
stellar ``halo'' in M31 \citep[e.g.,][]{dur04,irw05}.

%%%%%%%%%%%%%%%%%%%%%%%%%%%%%%%%%%%%%%%%%%%%%%%%%%%%%%%%%%%%%%%%%%%%%%%%%%%%%
\section{Conclusion}\label{sec:concl}
%%%%%%%%%%%%%%%%%%%%%%%%%%%%%%%%%%%%%%%%%%%%%%%%%%%%%%%%%%%%%%%%%%%%%%%%%%%%%
As part of a large spectroscopic survey of the outer halo of M31, we have
developed a method to isolate a clean sample of RGB stars in M31.  This
method uses five~diagnostics to isolate RGB stars: radial velocity, \ddo\
photometry, the EW of the \nai\ absorption feature, CMD position, and a
comparison of spectroscopic vs.\ photometric [Fe/H] estimates.  We also 
explored diagnostics based on the EWs of the K$_{7665}$ and K$_{7699}$
absorption features, and the strengths of the TiO bands at 7100, 7600, and
8500~\AA.  The first five diagnostics are more or less independent of one
another; the \ki\, \nai, and Ti0 diagnostics all show a dependence on stellar
effective temperature and are only effective for cool (red) stars.  The
overall likelihood based on a combination of diagnostics proves to be a much
more powerful tool than any of the individual diagnostics alone.  

The diagnostic method gives us the ability to confidently identify very
sparse populations of M31 RGB stars, and has led to the discovery of stars as
distant as $R\sim165$~kpc (in projection) from the galaxy's center.  This
technique has allowed us to reach surface brightness levels of
$\mu_V\approx35$~mag~arcsec$^{-2}$ and has enabled us to find an extended
stellar halo in M31 that follows an $R^{-2.6}$ surface brightness profile out
to $R\sim165$~kpc \citep{guh05}.  An investigation of the properties of these
RGB stars shows that the stars become increasingly metal-poor with increasing
distance from M31's center: beyond $R>60$~kpc, the mean metallicity of our
secure M31 RGB sample is found to be
$\langle\rm[Fe/H]_{phot}\rangle=-1.26\pm0.10$ \citep{kal06a}.  In addition to
identifying sparse populations, our method for isolating M31 RGB stars has
the benefit of permitting detailed dynamical studies of the M31 halo, since
the selection of stars does not depend on a radial velocity cut and the
resulting RGB sample is therefore not kinematically biased.  The application
of the diagnostic method to our M31 halo survey provides an unprecedented
level of sensitivity, allowing us to reliably study the extent, structure,
and kinematics of the remote M31 halo.

\acknowledgments 
We are grateful to Sandy Faber and the DEIMOS team for building an
outstanding instrument and for extensive help and guidance during its first
observing season.  We thank Peter Stetson, Jim Hesser, and James Clem for
help with the acquisition and reduction of CFHT/MegaCam images, Phil Choi,
Alison Coil, Geroge Helou, Drew Phillips, and Greg Wirth for observing some
DEIMOS masks on our behalf, Drew Phillips and Marla Geha for help with
slitmask designs, Leo Girardi for providing an extensive grid of theoretical
stellar isochrones, Ricardo Schiavon for expert advice on spectral features,
Jeff Lewis, Bill Mason, and Matt Radovan for fabrication of slitmasks, and
the DEEP2 team for allowing us use of the {\tt spec1d}/{\tt zspec} software.
The {\tt spec2d} data reduction pipeline for DEIMOS was developed at UC
Berkeley with support from NSF grant AST-0071048.  This project was supported
by an NSF Graduate Fellowship (K.M.G.), NSF grants AST-0307966 and
AST-0507483 and NASA/STScI grants GO-10265.02 and GO-10134.02 (P.G., K.M.G.,
J.S.K., and C.L.), NSF grants AST-0307842 and
AST-0307851, NASA/JPL contract 1228235, the David and Lucile Packard
Foundation, and The F.~H.~Levinson Fund of the Peninsula Community Foundation
(S.R.M., J.C.O., and R.J.P.), NSF grant AST-0307931 (R.M.R. and D.B.R.), and
a UCM Fundaci\'on del Amo Fellowship (A.J.C.).  J.S.K. is supported by NASA
through Hubble Fellowship grant HF-01185.01-A, awarded by the Space Telescope
Science Institute, which is operated by the Association of Universities for
Research in Astronomy, Incorporated, under NASA contract NAS5-26555.
%
% XXX {\bf Any additions/changes to the above acknowledgements?}

%%%%%%%%%%%% REFERENCES %%%%%%%%%%%%%%%%%%%%

\clearpage

%%%%%%%%%%%%%%%%%%%%%%% FIGURES %%%%%%%%%%%%%%%%%%%%%%%%%%%%%

\clearpage

\clearpage

\clearpage

\clearpage

\clearpage

\clearpage

\clearpage

\clearpage

\clearpage

\clearpage

\clearpage

\clearpage

\clearpage

\clearpage

\clearpage

\clearpage

\clearpage

\clearpage

\clearpage

\end{document}